\documentclass[twocolumn]{aastex631}
\usepackage{xspace,graphicx,apjfonts}

\newcommand\epsEri{$\epsilon$\,Eri\xspace}
\newcommand\sigDra{$\sigma$\,Dra\xspace}
\newcommand\Psc{107\,Psc\xspace}
\newcommand\xray{\mbox{X-ray}\xspace}
\newcommand\Rocrit{\mathrm{Ro_{crit}}}
\newcommand\Rosun{\mathrm{Ro_\odot}}
\newcommand\meanRhk{\log\left<R'_{\rm HK}\right>}

\journalinfo{The Astrophysical Journal (accepted version)}
\shorttitle{Magnetic Braking in K-type Stars}
\shortauthors{Metcalfe et al.}

\begin{document}

\title{\Large Testing the Rossby Paradigm: Weakened Magnetic Braking in early K-type Stars}

\author[0000-0003-4034-0416]{Travis S.~Metcalfe}%
\affiliation{Center for Solar-Stellar Connections, White Dwarf Research Corporation, 9020 Brumm Trail, Golden, CO 80403, USA}

\author[0000-0001-7624-9222]{Pascal Petit}%
\affiliation{Universit\'e de Toulouse, CNRS, CNES, 14 avenue Edouard Belin, 31400, Toulouse, France}

\author[0000-0002-4284-8638]{Jennifer L.~van~Saders}%
\affiliation{Institute for Astronomy, University of Hawai`i, 2680 Woodlawn Drive, Honolulu, HI 96822, USA}

\author[0000-0002-1242-5124]{Thomas R.~Ayres}%
\affiliation{Center for Astrophysics and Space Astronomy, 389 UCB, University of Colorado, Boulder, CO 80309, USA}

\author[0000-0002-1988-143X]{Derek Buzasi}%
\affiliation{Department of Astronomy \& Astrophysics, University of Chicago, Chicago, IL 60637, USA}

\author[0000-0003-3061-4591]{Oleg Kochukhov}%
\affiliation{Department of Physics and Astronomy, Uppsala University, Box 516, SE-75120 Uppsala, Sweden}

\author[0000-0002-3481-9052]{Keivan G.~Stassun}%
\affiliation{Vanderbilt University, Department of Physics \& Astronomy, 6301 Stevenson Center Lane, Nashville, TN 37235, USA}

\author[0000-0002-7549-7766]{Marc H.~Pinsonneault}%
\affiliation{Department of Astronomy, The Ohio State University, 140 West 18th Avenue, Columbus, OH 43210, USA}

\author[0000-0002-0551-046X]{Ilya V.~Ilyin}%
\affiliation{Leibniz-Institut f\"ur Astrophysik Potsdam (AIP), An der Sternwarte 16, D-14482 Potsdam, Germany}

\author[0000-0002-6192-6494]{Klaus G.~Strassmeier}%
\affiliation{Leibniz-Institut f\"ur Astrophysik Potsdam (AIP), An der Sternwarte 16, D-14482 Potsdam, Germany}

\author[0000-0002-3020-9409]{Adam J.~Finley}%
\affiliation{Universit\'e Paris-Saclay, Universit\'e Paris Cit\'e, CEA, CNRS, AIM, 91191, Gif-sur-Yvette, France}

\author[0000-0002-8854-3776]{Rafael A.~Garc\'ia}%
\affiliation{Universit\'e Paris-Saclay, Universit\'e Paris Cit\'e, CEA, CNRS, AIM, 91191, Gif-sur-Yvette, France}

\author[0000-0001-8832-4488]{Daniel Huber}%
\affiliation{Institute for Astronomy, University of Hawai`i, 2680 Woodlawn Drive, Honolulu, HI 96822, USA}
\affiliation{Sydney Institute for Astronomy (SIfA), School of Physics, University of Sydney, Camperdown, NSW 2006, Australia}

\author[0000-0003-4769-3273]{Yuxi (Lucy) Lu}%
\affiliation{Department of Astronomy, The Ohio State University, 140 West 18th Avenue, Columbus, OH 43210, USA}

\author[0000-0001-5986-3423]{Victor See}%
\affiliation{School of Physics \& Astronomy, University of Birmingham, Edgbaston, Birmingham B15 2TT, UK}

\begin{abstract}

There is an intricate relationship between the organization of large-scale magnetic 
fields by a stellar dynamo and the rate of angular momentum loss due to magnetized 
stellar winds. An essential ingredient for the operation of a large-scale dynamo is the 
Coriolis force, which imprints organizing flows on the global convective patterns and 
inhibits the complete cancellation of bipolar magnetic regions. Consequently, it is 
natural to expect a rotational threshold for large-scale dynamo action and for the 
efficient angular momentum loss that it mediates through magnetic braking. Here we 
present new observational constraints on magnetic braking for an evolutionary sequence of 
six early K-type stars. To determine the wind braking torque for each of our targets, we 
combine spectropolarimetric constraints on the large-scale magnetic field, Ly$\alpha$ or 
\xray constraints on the mass-loss rate, as well as uniform estimates of the stellar 
rotation period, mass, and radius. As identified previously from similar observations of 
hotter stars, we find that the wind braking torque decreases abruptly by more than an 
order of magnitude at a critical value of the stellar Rossby number. Given that all of 
the stars in our sample exhibit clear activity cycles, we suggest that weakened magnetic 
braking may coincide with the operation of a subcritical stellar dynamo.

\end{abstract}


\section{Introduction}\label{sec1}

Recent observations and numerical simulations suggest that magnetic stellar evolution is 
substantially more complex than has previously been assumed. The global stellar dynamo is 
responsible for the production and large-scale organization of magnetic fields, but the 
dominant scale of the magnetic morphology may change on stellar evolutionary timescales 
\citep{Buzasi1997, Garraffo2016, Garraffo2018}. It has been suggested that the dynamo 
might shift from a mode that weakly couples to the stellar wind in the saturated regime 
to a mode that strongly couples to it in the unsaturated regime \citep{Brown2014}. 
Subsequently, the dipole-dominated fields strongly couple the evolution of rotation and 
magnetism through angular momentum loss driven by magnetized stellar winds, a process 
known as magnetic braking \citep{WeberDavis1967, Skumanich1972, Kawaler1988}. When 
rotation eventually becomes too slow to imprint substantial Coriolis forces onto the 
global convective patterns, organizing flows such as differential rotation and meridional 
circulation may become weakened and the field might lose its large-scale organization, 
leading to a more complex morphology and decoupling the continued evolution of rotation 
and magnetism. These transitions appear to be accompanied by corresponding changes in 
magnetic variability---from multiperiodic or stochastic, to well-ordered periodic 
cycling, to constant or flat activity \citep{Metcalfe2017, Brun2022}.

The evidence for weakened magnetic braking (WMB) in old solar-type stars has expanded and 
solidified since it was initially suggested to explain anomalously rapid rotation in old 
field stars observed by the Kepler mission \citep{vanSaders2016}. The original sample 
included only 21 stars with asteroseismic ages \citep{Metcalfe2014} and rotation periods 
determined from spot modulations \citep{Garcia2014}. However, the imprint of WMB was also 
evident in the distribution of 34,000 rotation periods in the Kepler field 
\citep{McQuillan2014}. Subsequent forward modeling of this sample suggested that standard 
spin-down models could not reproduce the observed long-period edge \citep{vanSaders2019}, 
and more precise effective temperatures revealed a mixed population along the edge with 
diverse ages spanning the second half of main-sequence lifetimes \citep{David2022}. The 
asteroseismic sample was also expanded, with rotation periods determined from mode 
splitting rather than spot modulation \citep{Hall2021}, and the paucity of old slow 
rotators was confirmed with $v\sin{i}$ measurements \citep{Masuda2022}.

Modeling this phenomenon is complex because stellar dynamos are not simple functions of 
global stellar properties. Historically it was difficult to measure stellar magnetic 
fields directly, and investigators used stellar activity diagnostics as proxies for field 
strength. In a landmark paper, \cite{Noyes1984} demonstrated that activity indicators 
scaled with the ratio of the rotation period to the convective overturn timescale, or 
Rossby number ($\mathrm{Ro} \equiv P_{\rm rot}/\tau_c$), across a wide range of stellar 
rotation periods and masses. Recent work has confirmed that this scaling appears to hold 
for starspot filling factor \citep{Cao2022} and Zeeman broadening measurements 
\citep{Reiners2022}. It is therefore reasonable to hypothesize that the breakdown of the 
dynamo might also scale with Ro, and that the onset of WMB might be identified with a 
critical Rossby number ($\Rocrit$). Updated asteroseismic modeling and hierarchical 
Bayesian analysis has recently established a precise estimate of $\Rocrit$ for the onset 
of WMB \citep{Saunders2024}.

More direct observational constraints on magnetic braking have gradually become available 
over the past several years for a small sample of bright main-sequence stars. 
Spectropolarimetric snapshots of the late F-type stars 88\,Leo and $\rho$\,CrB suggested 
a substantial shift in magnetic morphology across $\Rocrit$ \citep{Metcalfe2019} 
accompanied by a large change in the estimated wind braking torque \citep{Metcalfe2021}. 
An evolutionary sequence of solar analogs provided additional constraints from archival 
Zeeman-Doppler Imaging (ZDI) maps of HD\,76151 and 18\,Sco along with new high 
signal-to-noise (S/N) snapshots of 16\,Cyg\,A and 16\,Cyg\,B \citep{Metcalfe2022}, 
reinforcing the conclusions drawn from the hotter stars. The extension of this approach 
to the late G-type stars 61\,UMa and $\tau$\,Cet \citep{Metcalfe2023a} suggested that the 
wind braking torque must decrease dramatically at the same value of $\Rocrit$ determined 
from indirect constraints \citep[cf.][]{Metcalfe2024a, Saunders2024}.

In this paper, we provide new constraints on magnetic braking for an evolutionary 
sequence of six early K-type stars (\epsEri, \sigDra, \Psc, HD\,103095, HD\,219134, 
HD\,166620) from the analysis of two published and two unpublished ZDI maps, as well as 
two recent high S/N snapshots from the Large Binocular Telescope (LBT). Based on the 
Rossby paradigm, these stars should exhibit a change in dynamo behavior at longer 
rotation periods than their F- and G-type counterparts. In Section~\ref{sec2} we describe 
the observations that were used to estimate the wind braking torque, including 
spectropolarimetry to infer the large-scale magnetic field strength and morphology 
(\S\ref{sec2.1}), archival Ly$\alpha$ and \xray observations to estimate the mass-loss 
rate (\S\ref{sec2.2}), and analysis of the broadband spectral energy distribution (SED) 
to estimate the stellar radius and mass (\S\ref{sec2.3}). In Section~\ref{sec3} we 
combine these inputs with stellar rotation periods adopted from the literature to match 
the observed stellar properties with rotational evolution models that include WMB 
(\S\ref{sec3.1}) and to estimate the wind braking torque for each of our targets 
(\S\ref{sec3.2}). Finally, in Section~\ref{sec4} we summarize and discuss our results, 
concluding that WMB may coincide with the operation of a subcritical stellar dynamo.

\section{Observations}\label{sec2}

Below we describe the new and archival observations that are needed to estimate the wind 
braking torque for each of our targets using the prescription of \cite{FinleyMatt2018}. 
Such estimates depend primarily on the large-scale magnetic field strength and morphology 
inferred from spectropolarimetry (\S\ref{sec2.1}), and the mass-loss rate inferred from 
Ly$\alpha$ observations and an empirical relation with \xray surface flux 
(\S\ref{sec2.2}). There are also relatively minor dependencies on stellar properties such 
as the mass and radius (\S\ref{sec2.3}).

\subsection{Spectropolarimetry}\label{sec2.1}

To constrain the strength and morphology of the large-scale magnetic field in each of our 
targets, we relied on new LBT snapshot observations of HD\,103095 and HD\,166620 
(\S\ref{sec2.1.1}), we constructed new ZDI maps from archival measurements of \sigDra 
(\S\ref{sec2.1.2}) and \Psc (\S\ref{sec2.1.3}), and we adopted the results from 
previously published ZDI maps for \epsEri \citep{Jeffers2014} and HD\,219134 
\citep{Folsom2018b}. We selected the 2008 map for \epsEri, which samples the mean 
activity level (as does the 2016 map of HD\,219134). For the wind braking calculations, 
we made the conservative assumption that the snapshot Stokes~$V$ profiles can be 
attributed entirely to an axisymmetric dipole field, which maximizes the resulting torque 
estimate. For the new and previously published ZDI maps, we followed the procedures 
described in \cite{Metcalfe2022} to convert the total magnetic flux in a given spherical 
harmonic degree ($B_\ell$) into equivalent polar field strengths for the dipole, 
quadrupole, and octupole components ($B_{\rm d}, B_{\rm q}, B_{\rm o}$), which are the 
required inputs for the wind braking prescription of \cite{FinleyMatt2018}.

\subsubsection{LBT Snapshot Observations}\label{sec2.1.1}

We acquired circular polarization (Stokes $V$) observations of HD\,103095 and HD\,166620 
on the nights of 2023~December~6 and 2024~July~5, respectively, using the Potsdam Echelle 
Polarimetric and Spectroscopic Instrument \citep[PEPSI;][]{Strassmeier2015} at the 
$2\times8.4$~m LBT. The instrument was configured for a resolving power of $R=130,000$ 
and covered the wavelength intervals 475--540~nm and 623--743~nm. The observational data 
were reduced as described in \citet{Metcalfe2019}. Anticipating the extremely weak 
polarization signals for these two stars, we adopted exposure times of 7200--8640~s, 
resulting in a peak S/N$\sim$3300--4400 per pixel in the extracted spectra.

 \begin{figure}[t!]
 \centering\includegraphics[width=\columnwidth]{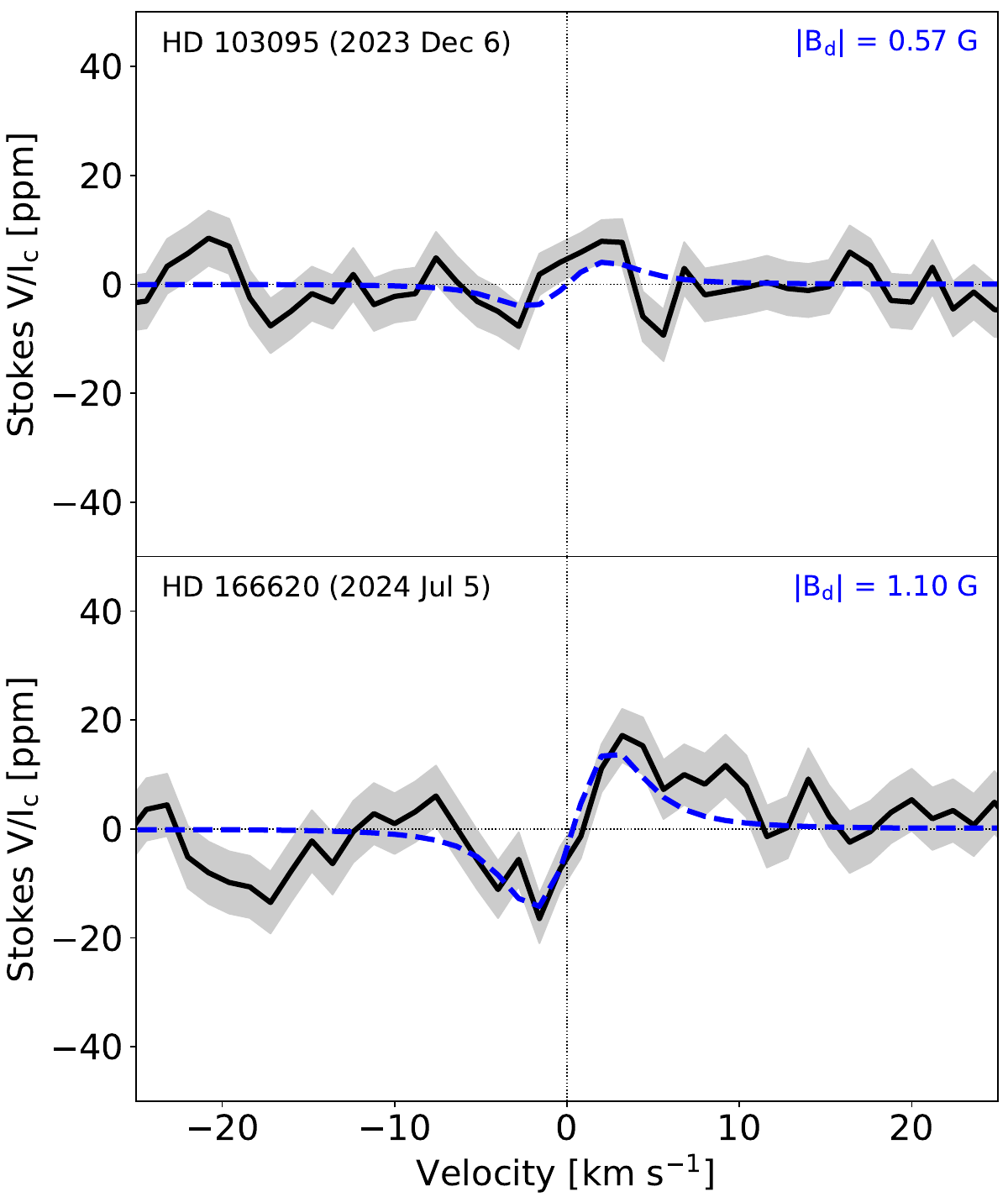}
 \caption{Stokes~$V$ polarization profiles for HD\,103095 (top) and HD\,166620 (bottom) 
from LBT observations on 2023~December~6 and 2024~July~5, respectively. The mean LSD 
profile is shown as a black line with uncertainties indicated by the gray shaded area. 
The dashed blue line is an axisymmetric model profile assuming dipole morphology with a 
fixed inclination.\label{fig1}}
 \end{figure}

The least-squares deconvolution \citep[LSD;][]{Kochukhov2010} technique was utilized to 
boost the S/N further by combining the profiles of all suitable metal lines. We 
constructed masks comprising 1630--2420 lines deeper than a few per cent of the continuum 
with the help of the VALD database \citep{Ryabchikova2015}, based on the spectroscopic 
parameters from \citet{ValentiFischer2005} for HD\,103095 and \citet{Brewer2016} for 
HD\,166620. The LSD analysis yielded mean Stokes $V$ profiles characterized by a 
polarimetric precision of about 5~ppm (see Figure~\ref{fig1}). This yielded a secure 
detection of the Zeeman polarization signature in HD\,166620, with a mean longitudinal 
magnetic field $\left<B_{\rm z}\right>=-0.28\pm0.07$~G (where the sign indicates the 
dominant field polarity). For HD\,103095 there was no significant detection, and the mean 
longitudinal field was consistent with zero, $\left<B_{\rm z}\right>=0.00\pm0.11$~G.

We applied the Stokes $V$ profile modeling technique described in \citet{Metcalfe2019} to 
constrain the strength of the global magnetic field by assuming an axisymmetric dipole 
morphology. We fixed the inclination of the stellar rotation axis using the analytic 
expressions from \cite{Bowler2023} to calculate posteriors given the measurements of $v 
\sin i$, rotation period, and radius. The posterior distribution peaked at $i=51^\circ$ 
for HD\,103095, yielding a best-fit dipole field strength of $B_{\rm d}=-0.57$, and 
$i=37^\circ$ for HD\,166620, yielding $B_{\rm d}=-1.10$~G. The predicted circular 
polarization profiles are illustrated with dashed blue lines in Figure~\ref{fig1}.

 \begin{figure*}[t!]
 \centering\includegraphics[height=4.125in]{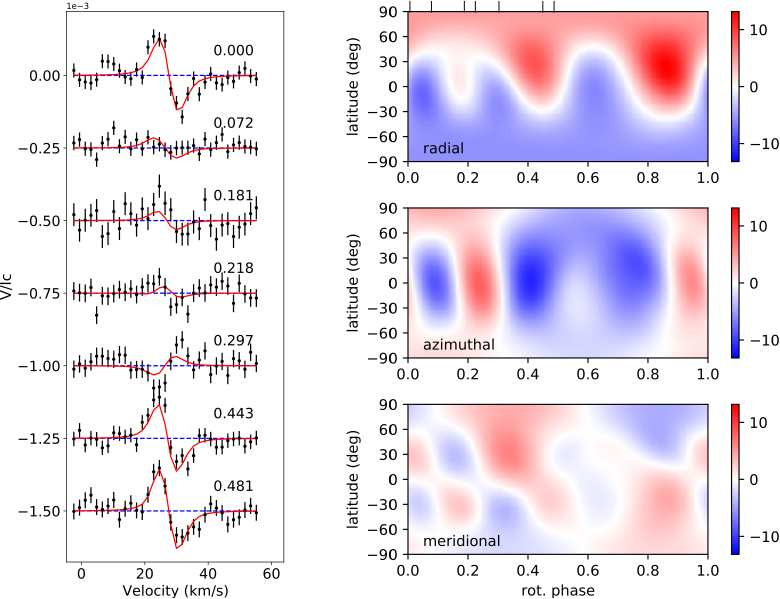} 
 \caption{Left: Stokes~$V$ LSD profiles obtained from NARVAL observations of \sigDra. 
Observations are shown as black dots (along with their error bars), while the ZDI model 
is displayed with red lines. Blue dashes show the null level, and the rotational phases 
of the observations are labeled on the right side of the plot. Successive observations 
are vertically shifted for clarity. Right: ZDI magnetic map of \sigDra in equirectangular 
projection. Each panel shows one component of the magnetic vector in spherical 
coordinates. The field strength is color-coded and expressed in Gauss. The rotation 
phases of observation are shown with vertical ticks above the top panel.\label{fig2}}
 \end{figure*}

\subsubsection{ZDI of \texorpdfstring{\sigDra}{sigma~Dra}}\label{sec2.1.2}

Multiple spectropolarimetric observations of \sigDra are available on the PolarBase 
archive \citep{Petit2014}. All of them were collected with the NARVAL spectropolarimeter 
\citep{Auriere2003}, offering a simultaneous recording of the full spectral domain 
between 370~nm and 1000~nm except a few, small wavelength intervals in the reddest part 
of the spectra. Following the standard data reduction performed with the 
\textsc{LibreEsprit} automated package \citep{Donati1997}, all observations were 
processed with the LSD multi-line method from which we extracted, for every spectrum, a 
single pseudo-line profile with greatly increased S/N. This standard approach 
\citep{Donati1997, Kochukhov2010} is a necessary step in the search for weak stellar 
magnetic fields, which generally produce polarized Zeeman signatures well below the noise 
level in individual lines. Thanks to the large spectral span of NARVAL, more than 5000 
photospheric spectral lines were used together, with a line list provided by the VALD 
database \citep{Ryabchikova2015}, for an effective temperature and surface gravity close 
to those of \sigDra, keeping only lines deeper than 40\% of the continuum level, and 
skipping wavelength intervals plagued by telluric bands or blended with broad 
chromospheric lines. The pseudo-line profiles have a normalized wavelength of 650~nm and 
a normalized Land\'e factor close to 1.2. For our dataset, the outcome of this procedure 
was the successful detection of Zeeman signatures in circular polarization (Stokes~$V$) 
for most of the available observations (as illustrated in Figure~\ref{fig2}), thanks to a 
final S/N close to 40,000.

Observations of \sigDra were gathered over three distinct epochs in 2007, 2009, and 2019. 
The dataset obtained in 2009 is the only one with a sampling of the stellar rotation 
sufficient for tomographic inversion. Our magnetic mapping was, therefore, focused on 
this specific time series. A quick analysis of the remaining data (not shown here) 
reveals that in 2007 the field polarity was always negative whenever a signature was 
detected (3 detected magnetic signatures, among 4 observations spanning 12 days). In 
2009, over a campaign spanning 13 days, 6 visits showed a positive polarity, one 
displayed a negative polarity, and one visit did not lead to a detection. In 2019, 3 
observations spanning 19 days consistently showed a negative polarity. The consistency of 
field polarities observed at each epoch, considered with the successive sign switches 
between epochs, may be indicative of a solar-like cycle taking place in \sigDra, with 
regular polarity flips of the global magnetic field \citep[e.g.,][]{BoroSaikia2018, 
doNascimento2023}. Indeed, the time series of chromospheric activity measurements 
presented in \cite{Baum2022} show that \sigDra has a 6.2 year activity cycle which was 
maximum in 2009 and near a magnetic minimum in 2007 and 2019, consistent with the 
observed polarity flips.

 \begin{figure*}[t!]
 \centering\includegraphics[height=4.125in]{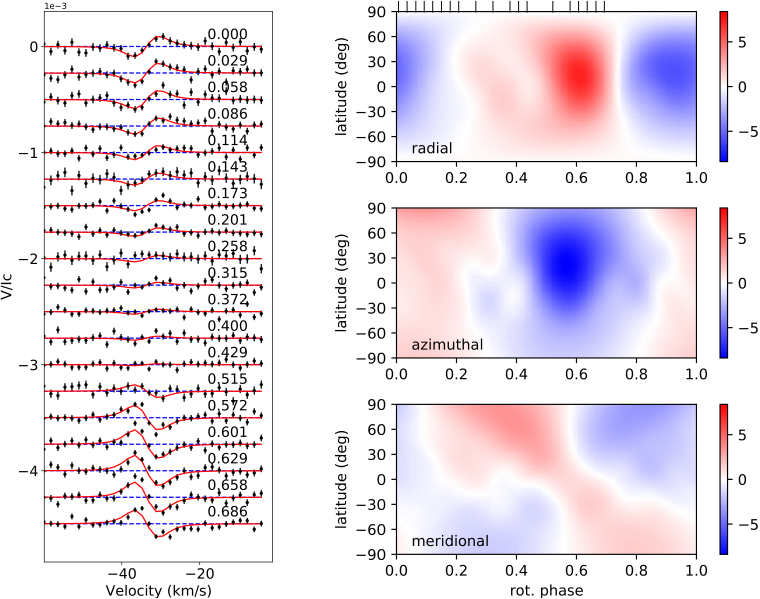}
 \caption{Left: Stokes~$V$ LSD profiles obtained from NARVAL observations of \Psc. 
Observations are shown as black dots (along with their error bars), while the ZDI model 
is displayed with red lines. Blue dashes show the null level, and the rotational phases 
of the observations are labeled on the right side of the plot. Successive observations 
are vertically shifted for clarity. Right: ZDI magnetic map of \Psc in equirectangular 
projection. Each panel shows one component of the magnetic vector in spherical 
coordinates. The field strength is color-coded and expressed in Gauss. The rotation 
phases of observation are shown with vertical ticks above the top panel.\label{fig3}}
 \end{figure*}

We used the series of 7 observations collected in 2009, between June 23 and July 6, to 
model the large-scale magnetic geometry of \sigDra using the ZDI method 
\citep{Semel1989}. We employed the Python code developed by \cite{Folsom2018a, 
Folsom2018b}, following the algorithm described by \cite{Donati2006}. We reproduced the 
procedure of \cite{Petit2021} to model the time series of LSD pseudo-line profiles, 
assuming that the observed variability of polarized signatures is induced entirely by 
rotational modulation. The ephemeris adopted to compute rotational phases was set as 
$HJD_{\rm obs} = HJD_{\rm 0} + P_{\rm rot} \times \Phi$ where $\Phi$ is the rotational 
phase, $HJD_{\rm obs}$ is the heliocentric Julian date of observation, $HJD_{\rm 0}$ is a 
reference date that we set to 2455005.61704, and $P_{\rm rot}$ is the rotation period, 
set to 27~d \citep{Baliunas1996}. The phase coverage obtained here presents a relatively 
dense sampling of phases between 0.0 and 0.5 but with a gap for the other half of the 
rotation period. Other standard input parameters of ZDI include a projected rotational 
velocity $v \sin i = 1.4$~km~s$^{-1}$ \citep{Brewer2016}, an angle between the spin axis 
and the line of sight set to 60$^\circ$, and a radial velocity equal to 
26.55~km~s$^{-1}$. Differential rotation was not included in the model because it cannot 
be reliably constrained with the available phase coverage. The spherical harmonic 
expansion of the magnetic field was limited to $\ell_{\rm max} = 10$ because no 
improvement of the fit was observed when a more complex field was allowed. A unit 
$\chi^2$ was obtained for the best model, showing that the assumptions listed above 
enabled us to produce a model fitting the data within their error bars. The outcome of 
the fit is illustrated with red lines in the left panel of Figure~\ref{fig2}, with the 
ZDI map shown in the right panel.

The magnetic geometry features a mean field strength of 7~G and a peak strength of 15~G. 
A majority of the magnetic energy (82\%) is reconstructed in the poloidal component. An 
inclined dipole stores about 45\% of the poloidal magnetic energy, with a polar strength 
of 7~G. The field configuration is mostly non-axisymmetric, with only 35\% of the 
magnetic energy in $m = 0$ spherical harmonic modes.

\subsubsection{ZDI of \texorpdfstring{\Psc}{107~Psc}}\label{sec2.1.3}

The spectropolarimetric dataset modeled for \Psc was obtained in 2008, with 19 visits 
between January 22 and February 15 using the NARVAL spectropolarimeter. Two other 
observing epochs are available on the PolarBase archive---one from early 2007, and a 
second from the summer of 2007. We did not focus on these earlier datasets, because they 
displayed a sparser phase coverage (early 2007) or suffered from a lower detection rate 
(mid-2007). The procedures applied to extract Zeeman signatures and to model the magnetic 
geometry follow the same steps as those described above for \sigDra.

The set of Stokes~$V$ LSD signatures obtained in 2008 are shown in Figure~\ref{fig3}. 
They display a negative polarity until January 29. After this date, the signatures fade 
away until February 9, when a positive polarity progressively grows. Unfortunately, the 
time span available for this run was limited to 24 days, preventing us from observing 
about 30\% of the complete rotation to recover the negative polarity. The simple 
observation of an alternation of polarities tells us, before any tomographic modeling, 
that the magnetic geometry at this epoch was very non-axisymmetric. We note that the 
observations in 2007 are mostly consistent with this picture, with both polarities 
successively observed in early 2007 and an amplitude compatible with 2008 values, while 
the few detections obtained during the summer display a positive polarity, again of a 
roughly similar amplitude.

The ZDI map was computed assuming a rotation period of 35~d, $v \sin i = 0.1$~km~s$^{-1}$ 
\citep{Brewer2016}, a mean radial velocity of $-33.7$~km~s$^{-1}$, and a spin axis 
inclination of 60$^\circ$. The reference rotational phase was set to Julian date 
$HJD_{\rm 0} = 2454488.28715$. As for \sigDra, the level of detail in the Stokes~$V$ data 
did not require a spherical harmonic expansion above $\ell_{\rm max} = 10$. This set of 
parameters allowed the ZDI inversion to reach a reduced $\chi^2$ of 0.98. The result of 
the fit is shown with the Stokes~$V$ profiles in the left panel of Figure~\ref{fig3}, 
with the ZDI map shown in the right panel.

\begin{deluxetable*}{@{}lccccccc}
  \setlength{\tabcolsep}{9pt}
  \tabletypesize{\footnotesize}
  \tablecaption{Stellar Properties of the Evolutionary Sequence\label{tab1}}
  \tablehead{\colhead{}              & \colhead{\epsEri}         & \colhead{\sigDra}         & \colhead{\Psc}            &
             \colhead{HD\,103095}    & \colhead{HD\,219134}      & \colhead{HD\,166620}      & \colhead{Sources}         }
  \startdata
$T_{\rm eff}$ (K)                    & $5146 \pm 44$             & $5242 \pm 25$             & $5190 \pm 25$             &
             $4950 \pm 44$           & $4835 \pm 44$             & $4970 \pm 25$             & 1, 2                      \\
$[$M/H$]$ (dex)                      & $0.00 \pm 0.03$           & $-0.21 \pm 0.01$          & $-0.03 \pm 0.01$          &
             $-1.16 \pm 0.03$        & $+0.09 \pm 0.03$          & $-0.10 \pm 0.01$          & 1, 2                      \\
$\log g$ (dex)                       & $4.57 \pm 0.06$           & $4.56 \pm 0.03$           & $4.51 \pm 0.03$           &
             $4.65 \pm 0.06$         & $4.56 \pm 0.06$           & $4.51 \pm 0.03$           & 1, 2                      \\
$P_{\rm cyc}$ (yr)                   & $2.95 \pm 0.03$           & $6.2 \pm 0.1$             & $9.6 \pm 0.1$             &
             $7.30 \pm 0.08$         & $11.6 \pm 0.3$            & $15.8 \pm 0.3$            & 3, 4, 5, 6                \\
$\meanRhk$ (dex)                     & $-4.455$                  & $-4.832$                  & $-4.912$                  &
             $-4.896$                & $-4.890$                  & $-4.955$                  & 6, 7                      \\
$P_{\rm rot}$ (days)                 & $12 \pm 0.5$              & $27 \pm 0.5$              & $35 \pm 0.5$              &
             $31\pm 0.5$             & $42 \pm 0.9$              & $43\pm 0.5$               & 7, 8                      \\
Inclination ($^\circ$)               & $46 \pm 2$                & $60$                      & $60$                      &
             $51$                    & $77 \pm 8$                & $37$                      & 8, 9, 10                  \\
$|B_{\rm d}|$ (G)                    & $14.6$                    & $5.68$                    & $4.24$                    &
             $0.57 \pm 0.03$         & $2.39$                    & $1.10^{+0.42}_{-0.40}$    & 8, 9, 10                  \\
$|B_{\rm q}|$ (G)                    & $8.78$                    & $4.82$                    & $2.77$                    &
             $\cdots$                & $4.05$                    & $\cdots$                  & 8, 9, 10                  \\
$|B_{\rm o}|$ (G)                    & $5.90$                    & $4.76$                    & $1.37$                    &
             $\cdots$                & $1.19$                    & $\cdots$                  & 8, 9, 10                  \\
$L_{\rm X}$ ($10^{27}$~erg~s$^{-1}$) & $19.1 \pm 3.5$            & $5.1^{+1.9}_{-1.4}$       & $1.2^{+0.6}_{-0.4}$       &
             $0.17^{+0.05}_{-0.04}$  & $0.72 \pm 0.05$           & $1.1 \pm 0.3$             & 11                        \\
Mass-loss rate ($\dot{M}_\odot$)     & $30^{+30}_{-15}$          & $3.05^{+3.27}_{-1.72}$    & $0.93^{+1.00}_{-0.54}$    &
             $0.30^{+0.31}_{-0.17}$  & $0.50^{+0.50}_{-0.25}$    & $0.92^{+0.94}_{-0.51}$    & 11, 12                    \\
Luminosity ($L_\odot$)               & $0.304 \pm 0.007$         & $0.402 \pm 0.009$         & $0.431 \pm 0.010$         &
             $0.222 \pm 0.008$       & $0.257 \pm 0.009$         & $0.336 \pm 0.008$         & 13                        \\
Radius ($R_\odot$)                   & $0.694 \pm 0.014$         & $0.772 \pm 0.005$         & $0.813 \pm 0.012$         &
             $0.641 \pm 0.017$       & $0.724 \pm 0.015$         & $0.782 \pm 0.012$         & 13, 14                    \\
Mass ($M_\odot$)                     & $0.86 \pm 0.05$           & $0.84 \pm 0.01$           & $0.87 \pm 0.05$           &
             $0.60 \pm 0.04$         & $0.80 \pm 0.05$           & $0.80 \pm 0.05$           & 13, 14                    \\
Age, astero (Gyr)                    & $\cdots$                  & $4.54 \pm 0.92$           & $\cdots$                  &
             $\cdots$                & $10.2 \pm 1.5$            & $\cdots$                  & 14, 15                    \\
Age, gyro (Gyr)                      & $1.3_{-0.20}^{+0.23}$     & $5.12_{-0.87}^{+0.94}$    & $6.88_{-1.13}^{+1.19}$    &
             $7.12_{-1.21}^{+1.34}$  & $8.18_{-1.40}^{+1.48}$    & $9.16_{-1.40}^{+1.47}$    & 16                        \\
Age, WMB (Gyr)                       & $1.46_{-0.22}^{+0.25}$    & $5.65_{-0.97}^{+1.14}$    & $7.72_{-1.35}^{+1.80}$    &
             $8.51_{-1.77}^{+2.58}$  & $8.83_{-1.52}^{+1.69}$    & $9.5_{-1.65}^{+2.15}$     & 16                        \\
%
%
Rossby number ($\Rosun$)             & $0.33 \pm 0.01$           & $0.78 \pm 0.02$           & $0.93 \pm 0.02$           &
             $0.90 \pm 0.02$         & $0.89 \pm 0.02$           & $1.03 \pm 0.02$           & 17                        \\
  \hline 
Torque ($10^{30}$~erg)               & $13.9^{+8.9}_{-5.4}$      & $1.06^{+0.58}_{-0.41}$    & $0.383^{+0.230}_{-0.161}$ &
           $0.019^{+0.014}_{-0.009}$ & $0.095^{+0.058}_{-0.036}$ & $0.081^{+0.091}_{-0.049}$ & 17                        \\
  \enddata
  \tablerefs{(1)~\cite{ValentiFischer2005}; (2)~\cite{Brewer2016}; (3)~\cite{Metcalfe2013}; (4)~\cite{Baum2022};
(5)~\cite{Baliunas1995}; (6)~\cite{Johnson2016}; (7)~\cite{Baliunas1996}; (8)~\cite{Folsom2018b}; (9)~\cite{Jeffers2014}; 
(10)~Section~\ref{sec2.1}; (11)~Section~\ref{sec2.2}; (12)~\cite{Wood2021}; (13)~Section~\ref{sec2.3}; (14)~\cite{Hon2024}; 
(15)~\cite{Li2025}; (16)~Section~\ref{sec3.1}; (17)~Section~\ref{sec3.2} }
\vspace*{-24pt}
\end{deluxetable*}
\vspace*{-24pt}

The magnetic geometry features an average field strength of 4~G. A majority (84\%) of the 
magnetic energy is reconstructed in the poloidal component, and a very tilted dipole 
accounts for slightly more than 80\% of the poloidal magnetic energy. As expected, the 
field configuration is very non-axisymmetric, with only 12\% of the magnetic energy in $m 
= 0$ spherical harmonic modes. We note that this same dataset has been used to 
reconstruct the ZDI map of \Psc for two earlier studies \citep{vidotto2014, see2016}. The 
updated model is mostly consistent with the previous version, although the field strength 
recovered here is slightly larger. This difference is primarily due to modified criteria 
adopted to ensure that the ZDI code does not overfit or underfit the data \citep[we 
follow the method employed by][]{bellotti2024}. Stellar properties are listed in 
Table~\ref{tab1}.

\subsection{X-Ray Data}\label{sec2.2}

Our target stars have a variety of archival \xray data, primarily from the venerable 
1990s R{\"o}ngtensatelit (ROSAT), both the all-sky survey (RASS, using the PSPC 
proportional counter) and post-survey pointings with the PSPC and HRI 
(micro-channel-based sensor). In some cases, there are contemporary measurements from 
XMM-Newton with its European Photon Imaging Camera (EPIC; suite of three CCDs: pn, MOS1, 
MOS2), and the Chandra \xray Observatory with its Advanced CCD Imaging Spectrometer 
(ACIS).

ROSAT count rates (CR) were taken from mission catalogs (rass2rxs for the RASS; rospspc 
for PSPC pointings; roshri for HRI pointings) hosted by the High-Energy Astrophysics 
Science Archive Research Center (HEASARC) at the NASA Goddard Space Flight Center. The 
RASS exposure times described below refer to the total of many brief scans during several 
days of sky coverage on the target, whereas the PSPC and HRI pointings were continuous 
integrations (modulo corrections for telemetry deadtime; likewise for the XMM and Chandra 
observations).

XMM and Chandra level-2 event lists (from HEASARC or the Chandra archive) were specially 
processed and measured according to protocols outlined by \cite{Ayres2024}. Counts inside 
an instrument-defined detection cell centered on the target were parsed into 
evenly-spaced time intervals, corrected for background, deadtime, and encircled energy 
factor. The time series was then subjected to a ``flare filter'' that removed 
percentages, possibly unequal, of the highest and lowest CR bins, according to a 
pre-defined elimination hierarchy. The ``Olympic'' filter suppressed transient 
enhancements (flares), as well as occasional dropouts (e.g., telemetry-saturating 
``background flares'' for XMM). This filtering was only relevant for our most active 
target (\epsEri), and had a negligible influence on the results (see \S\ref{sec2.2.1}). 
The adopted detection cells were 80\% encircled energy for XMM (cell radius varied 
between 20{\arcsec}--23{\arcsec} depending on EPIC module), and 95\% for the higher 
resolution Chandra ACIS ($r= 1{\farcs}5$). These cell sizes were designed to maximize the 
source counts, while minimizing unwanted background events, based on the specific 
instrument characteristics.

Filtered average CR for XMM and Chandra, and catalog CR for ROSAT, were converted to 
\xray fluxes at Earth using an optimization scheme based on a grid of coronal 
emission-measure models \citep[for details, see][]{Ayres2024}. The output energy range 
was 0.1--2.4~keV (``ROSAT standard band''), for the unabsorbed flux (i.e., corrected for 
interstellar absorption). ISM column densities, $N_{\rm H}$, for these nearby stars were 
set to a nominal $1{\times}10^{18}$~cm$^{-2}$. The adopted bandpass is well-suited to the 
spectral distributions of late-type coronal sources, and well-matched to the input energy 
ranges of the various \xray missions considered here. Descriptions of the individual 
stellar measurements are given below.

\subsubsection{X-Ray Luminosities}\label{sec2.2.1}

The K2 dwarf \epsEri (HD\,22049, $d=3.2$\,pc) has been observed 17 times by XMM, mainly 
between 2015 February and 2022 January, but with a single earlier pointing in 2003 
January. The exposures ranged from 7.6~ks to 21.5~ks, averaging 11~ks. Based on our new 
processing, the XMM time series had a mean $L_{\rm X}= (19.1 \pm 3.5) \times 
10^{27}$~erg~s$^{-1}$, representing a time-average over multiple short activity cycles 
(2015--2022). \cite{Coffaro2020} and \cite{Fuhrmeister2023} have previously described 
most of the XMM pointings. \citeauthor{Fuhrmeister2023} reported \xray luminosities for 
seven additional XMM exposures obtained after the nine published by 
\citeauthor{Coffaro2020}, but apparently missing one from 2021 February. Combining the 
two time series yields $L_{\rm X}= (19 \pm 4) \times 10^{27}$~erg~s$^{-1}$, the same as 
our new processing of the full dataset. However, the agreement is probably coincidental 
because the earlier result refers to the narrower 0.2--2.0~keV energy band, and without 
flare filtering. Their result would be higher for our broader 0.1--2.4~keV energy band, 
but lower if the relatively few \epsEri flares were removed.

The K0 dwarf \sigDra (HD\,185144, $d=5.8$\,pc) was observed several times in the ROSAT 
era: during the all-sky survey (1991 August, near cycle maximum) with a CR of 0.29~cps in 
1.5~ks (deeper than normal owing to the high ecliptic latitude of the star, favored by 
the RASS scanning pattern); PSPC pointings near cycle maxima in 1992 November (0.17~cps 
in 2.7~ks) and 1997 February (0.37~cps in 1.7~ks); and a brief HRI exposure in 1998 April 
near cycle minimum (0.036~cps in 0.9~ks). \xray luminosities derived from the various 
ROSAT measurements were in the range (3.3--8.0)$\times 10^{27}$~erg~s$^{-1}$. We adopt 
the logarithmic average, $L_{\rm X}=(5.1_{-1.4}^{+1.9}) \times 10^{27}$~erg~s$^{-1}$. The 
variations exceed those seen in the extensive time series of the more active \epsEri, but 
are comparable to the larger swings of lower activity stars like $\alpha$\,Cen\,B 
(HD\,128621, K1V) and 61\,Cyg\,A (HD\,201091, K5V), which have more exaggerated starspot 
cycles in \mbox{X-rays} \citep[see][]{Ayres2024}. The influence of flares on the several 
ROSAT pointings is unknown, although such outbursts tend to be rare among the 
low-activity K dwarfs compared to their more active cousins.

The K1 dwarf \Psc (HD\,10476, $d=7.6$\,pc) was not detected in the RASS, but there were 
two later ROSAT pointings: PSPC in 1993 July near cycle minimum (0.027~cps in 4.1~ks) and 
HRI in 1997 July near cycle maximum (0.0087~cps in 3.1~ks). The calibrated \xray 
luminosities differ by a factor of two: $0.79 \times 10^{27}$~erg~s$^{-1}$ for PSPC and 
$1.7 \times 10^{27}$~erg~s$^{-1}$ for HRI. We adopt the logarithmic average $L_{\rm 
X}=(1.2_{-0.4}^{+0.6}) \times 10^{27}$~erg~s$^{-1}$, similar to 61\,Cyg\,A 
\citep[e.g.,][]{Ayres2024}.

The K1 dwarf HD\,103095 ($d=9.2$\,pc) was not detected in the RASS, but there were two 
subsequent ROSAT pointings: PSPC in 1993 May near cycle maximum (0.0055~cps in 2.8~ks), 
and HRI in 1996 November near cycle minimum (no detection in 6.7~ks). Chandra obtained 
one observation:  32.8~ks with ACIS-S in 2009 February near the mean activity level. The 
measured CR was only 0.0010~cps (consistent with the HRI non-detection). The calibrated 
\xray luminosities of the ROSAT and Chandra detections differ by a factor of two: $0.21 
\times 10^{27}$~erg~s$^{-1}$ for PSPC and $0.13 \times 10^{27}$ for ACIS-S. We adopt the 
logarithmic average, $L_{\rm X}=(0.17_{-0.04}^{+0.05}) \times 10^{27}$~erg~s$^{-1}$.

The K3 dwarf HD\,219134 ($d=6.5$\,pc) was detected in the RASS (1991 January, near cycle 
minimum) with a CR of 0.032~cps in 0.5~ks. The implied $L_{\rm X}$ is $(0.7 \pm 0.2) 
\times 10^{27}$~erg~s$^{-1}$. There was also a 38~ks XMM pointing in 2016 June near the 
mean activity level. We adopt the XMM value (median of pn, MOS1, MOS2), $L_{\rm X}=(0.72 
\pm 0.05) \times 10^{27}$ erg s$^{-1}$, consistent with the RASS result but more precise 
due to the higher quality XMM instrumentation and the deeper exposure.

The K2 dwarf HD\,166620 ($d=11.1$\,pc) was not detected in the RASS, but there was a 
later HRI pointing in 1996 October near cycle maximum (0.0026~cps in 6.3~ks). The HRI CR 
implies $L_{\rm X}= (1.1 \pm 0.3) \times 10^{27}$~erg~s$^{-1}$, which we adopt for our 
analysis.

 \begin{figure*}[t!]
 \centering\includegraphics[width=\textwidth]{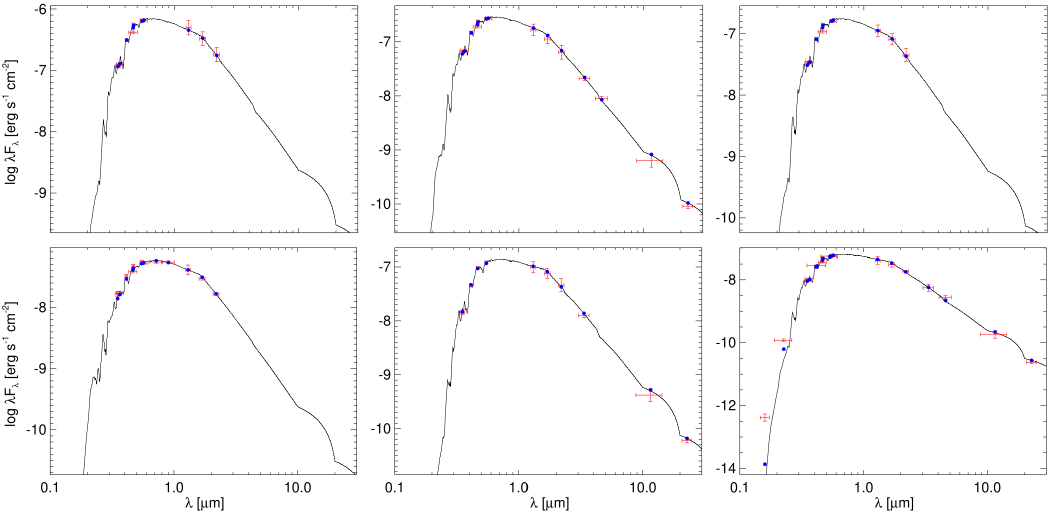}
 \caption{Spectral energy distributions of (top, left to right) \epsEri, \sigDra, and 
\Psc, and (bottom) HD\,103095, HD\,219134, and HD\,166620. Red symbols represent the 
observed photometric measurements, and the horizontal bars represent the effective width 
of the bandpass. Blue symbols are the model fluxes from the best-fit Kurucz atmosphere 
model (black).\label{fig4}}
 \end{figure*}

\subsubsection{Mass-Loss Rates}\label{sec2.2.2}

Wind braking torques depend on the mass-loss rate and the Alfv\'en radius. Mass-loss 
rates are not directly measured in most of our targets, so we use the coronal heating 
rate (as inferred from \xray luminosities) to estimate the mass-loss rates. Combining the 
cycle averaged \xray luminosities adopted above and the SED radii determined below 
(Section~\ref{sec2.3}), we calculated \xray surface fluxes ($F_{\rm X}$) for each of our 
targets. \cite{Wood2021} established an empirical relation between $F_{\rm X}$ and the 
mass-loss rate per unit surface area, $\dot{M} \propto F_{\rm X}^{0.77 \pm 0.04}$, which 
agrees well with the predictions of \cite{Suzuki2013}. \cite{Wood2018} found a steeper 
relation by fitting only the data for GK dwarfs ($\dot{M} \propto F_{\rm X}^{1.29}$), but 
the relation from \cite{Wood2021} is the more conservative choice because it predicts a 
slower decline of $\dot{M}$ at low $F_{\rm X}$.

For two of our targets (\epsEri and HD\,219134 = GJ\,892) we adopt the mass-loss rates 
inferred directly from Ly$\alpha$ measurements, as tabulated in \cite{Wood2021}. Although 
no formal uncertainties are given in that paper, the authors note: ``We have in the past 
estimated that $\dot{M}$ values measured in this way should be accurate to within about a 
factor of 2 \citep{Wood2005a}, with important systematic uncertainties including the 
unknown degree of variation in ISM properties from star to star and possible differences 
in stellar wind speed.'' These systematic errors dominate the total uncertainty of our 
estimated mass-loss rates, so we combine the quoted factor of two error in quadrature 
with the statistical uncertainty for the mass-loss rates listed in Table~\ref{tab1}.

\subsection{Stellar Properties}\label{sec2.3}

To obtain empirical constraints on the stellar luminosities, radii, and masses, we 
performed an analysis of the broadband SED of each star together with the Gaia DR3 
parallaxes, following the procedures described in \cite{Stassun2016} and 
\cite{Stassun2017, Stassun2018}. No systematic offset to the parallaxes was applied 
\citep[see, e.g.,][]{StassunTorres2021}. Depending on the available published broadband 
photometry for each star, we adopted some combination of the $JHK_S$ magnitudes from 
2MASS, the W1--W4 magnitudes from WISE, the $UBV$ magnitudes from \cite{Mermilliod2006}, 
the Str\"omgren $ubvy$ magnitudes from \cite{Paunzen2015}, and the FUV and NUV magnitudes 
from GALEX. Together, the available photometry spans the full stellar SED over at least 
the wavelength range 0.4--10~$\mu$m in most cases, and in some cases as large as 
0.2--20~$\mu$m (see Figure~\ref{fig4}).

We performed fits using Kurucz stellar atmosphere models, with the effective temperature 
$T_{\rm eff}$, surface gravity $\log g$ and metallicity [M/H] from the published 
spectroscopic analyses (see Table~\ref{tab1}). The extinction $A_V$ was fixed at zero in 
all cases due to the very close proximity of the stars. The resulting fits have a reduced 
$\chi^2 \approx 1$ in all cases. Integrating the model SED gives the bolometric flux at 
Earth, $F_{\rm bol}$. Taking $F_{\rm bol}$ together with the Gaia parallax 
\citep{Gaia2021} yields the bolometric luminosity $L_{\rm bol}$. The stellar radius then 
follows directly from the Stefan-Boltzmann relation. Finally, we estimate the stellar 
mass from the empirical eclipsing-binary based relations of \citet{Torres2010}. The 
results are listed in Table~\ref{tab1}.

\section{Modeling}\label{sec3}

Below we use rotational evolution models to match the observational constraints described 
above, establishing the chronology of our targets in terms of stellar age 
(\S\ref{sec3.1}). We then use the prescription of \cite{FinleyMatt2018} to estimate the 
wind braking torque for each star in the evolutionary sequence by combining the inputs 
from Section~\ref{sec2} with published rotation periods (\S\ref{sec3.2}). The final 
result is an observationally constrained picture of how magnetic braking changes with 
Rossby number on stellar evolutionary timescales, which we summarize and discuss in 
Section~\ref{sec4}.

\subsection{Gyrochronology}\label{sec3.1}

We modeled the rotational evolution of our targets with an approach similar to that 
described in \citet{Metcalfe2023a, Metcalfe2023b, Metcalfe2024a, Metcalfe2024b} with some 
minor amendments. We considered both a ``standard spin-down'' case with 
\mbox{$\dot{M}\sim L_{\rm bol}/\mathrm{Ro}^2$} and \mbox{$B \sim P_{\rm 
phot}^{1/2}/\mathrm{Ro}$} (where $P_{\rm phot}$ is the photospheric pressure), and a WMB 
case where magnetic braking ceases above $\Rocrit$ while angular momentum is conserved.

Our model grids are constructed with the Yale Rotating Evolution Code \citep[YREC, 
see][]{Pinsonneault1989} and are identical to those described in \citet{Metcalfe2020} and 
\citet{vanSaders2013} with two notable modifications: first, we adopt the model 
atmospheres of \citet{Kurucz1993} as boundary conditions, and second we include helium 
and heavy element diffusion and gravitational settling via the prescription of 
\citet{Thoul1994}. In previous papers \citep{Metcalfe2023a, Metcalfe2023b, Metcalfe2024a, 
Metcalfe2024b, Saunders2024} we adopted an Eddington atmosphere over atmospheric tables 
to better match the asteroseismic modeling, which requires a detailed atmospheric 
structure. However, our target K dwarfs mostly lack asteroseismic detections (see 
Appendix~\ref{appa}), and are sufficiently cool that a gray Eddington atmosphere is a 
particularly poor model. Because our target stars are both old and slowly rotating, we 
also opt to include diffusion and gravitational settling, as these processes can 
significantly impact the relationship between the observed surface metallicity and 
inferred bulk starting metallicity \citep{Dotter2017}. A solar calibration yields a solar 
bulk helium mass fraction of 0.2744, a mixing length of 1.967 pressure scale heights, and 
default \citet{Thoul1994} gravitational diffusion coefficients multiplied by 0.7326 to 
match the solar envelope helium abundance \citep{Basu1995}. We adopt a chemical 
enrichment law of the form $Y = Y_p + \frac{dY}{dZ} Z$, with the primordial helium $Y_p = 
0.249$ and $dY/dZ= (Y_{\odot}-Y_p)/Z_{\odot}=1.352$. A solar model rotating at 25.4 days 
has a Rossby number in this model grid of $\Rosun = 2.36$.

In the absence of tight asteroseismic constraints for most targets, we used the 
rotational evolution model of \cite{Saunders2024} coupled with the constraints on radius, 
effective temperature, surface metallicity, and rotation period to infer the stellar 
masses, ages, and Rossby numbers. This braking law includes the effects of WMB, but does 
not attempt to model the ``stalled spin-down" observed in intermediate-age K dwarfs 
thought to be due to an epoch of internal angular momentum transport \citep{Curtis2019, 
Curtis2020, Spada2020}. Neglecting this feature will tend to bias our model ages at 
$\sim$1--2~Gyr, in the sense that our ages should appear younger than those inferred from 
models that include the stalled spin-down. The effect on old stars is modest, and in all 
cases our approach should preserve the rank-ordering of systems in age and Rossby number.

We adopted the observational constraints and their associated uncertainties from 
Table~\ref{tab1}. We inflated the reported random errors with systematic error estimates 
added in quadrature: 2\% in $T_{\rm eff}$ and 4.2\% in radius following the 
recommendation of \citet{Tayar2022}, 0.1~dex in surface abundance, and 10\% in period 
\citep{Epstein2014}. We placed broad Gaussian priors (specified as a central value and a 
1$\sigma$ width) on the stellar masses ($0.8\pm0.2\ M_\odot$), bulk abundance 
($\pm0.5$~dex centered on the observed surface abundance), age ($4.5 \pm 5$~Gyr) and 
mixing length $\alpha$ ($1.5 \pm 0.5$) for all target stars. We adopted the braking law 
parameters from \citet{Saunders2024}, appropriately re-normalized to reflect the slightly 
different solar convective overturn timescale and photospheric pressure in the new model 
grid.

We validated our modeling procedure against other age inference techniques. We inferred 
rotation-based ages for three asteroseismic targets of similar surface temperatures with 
precise ages from \citet{Saunders2024}---KIC\,11772920, KIC\,9025370, and KIC\,7970740. 
Our non-seismic, rotation-based ages using the WMB model agree with the quoted 
asteroseismic ages within $0.8\sigma$, $1.2\sigma$, and $0.8\sigma$, respectively, with 
the caveat that these stars were themselves utilized in the \citet{Saunders2024} 
calibration process. For the two K dwarfs in our sample with asteroseismic ages (which 
were not used as priors in the fit), our WMB rotation-based ages agree with the 
asteroseismic values within $1\sigma$. Cool K dwarfs evolve slowly, and therefore cannot 
be age-dated with standard isochrone methods. Isochrone fitting for our target stars with 
the YREC model grid yielded uninformative ages, as expected given their positions near 
the main-sequence.

A gyrochronology model \texttt{GPgyro} \citep{Lu2024} calibrated on gyro-kinematic ages 
\citep{Lu2021} with rotation measurements from Kepler \citep{McQuillan2014, Santos2019}, 
and ZTF \citep{Lu2022} was also used to infer ages for the targets. Gyro-kinematic ages 
use samples of stars selected to have similar rotation periods, temperature, and absolute 
Gaia $G$ magnitude to define mono-age kinematic sets, instead of the traditional 
selection in physical space. As such, the method performs well in populations where the 
spin-down is simple, but may provide biased ages when stars of different ages can have 
similar rotation periods---such as objects undergoing WMB or stalled spin-down. Our WMB 
ages agree within $1\sigma$ for \Psc, HD\,219134, and HD\,166620. They agree within 
$2.2\sigma$ for \epsEri, $1.7\sigma$ for \sigDra, and $1.8\sigma$ for HD\,103095. 
Gyro-kinematic ages are known to have issues for stars as young as \epsEri, and do not 
explicitly account for metallicity, which may help to explain \sigDra and HD\,103095. 
Agreement in the old, solar metallicity targets is excellent.

Model ages are provided in Table~\ref{tab1} for both the assumption of standard spin-down 
(``Age, gyro'') and weakened magnetic braking (``Age, WMB''). As expected for a star with 
Ro\,$<\Rocrit$, both values for \sigDra are consistent with the asteroseismic age from 
\cite{Hon2024}. The asteroseismic age for HD\,219134 \citep{Li2025} agrees with the WMB 
age, but is slightly older than the standard spin-down age.

\subsection{Wind Braking Torque}\label{sec3.2}

\begin{figure}[t!]
 \centering\includegraphics[width=\columnwidth]{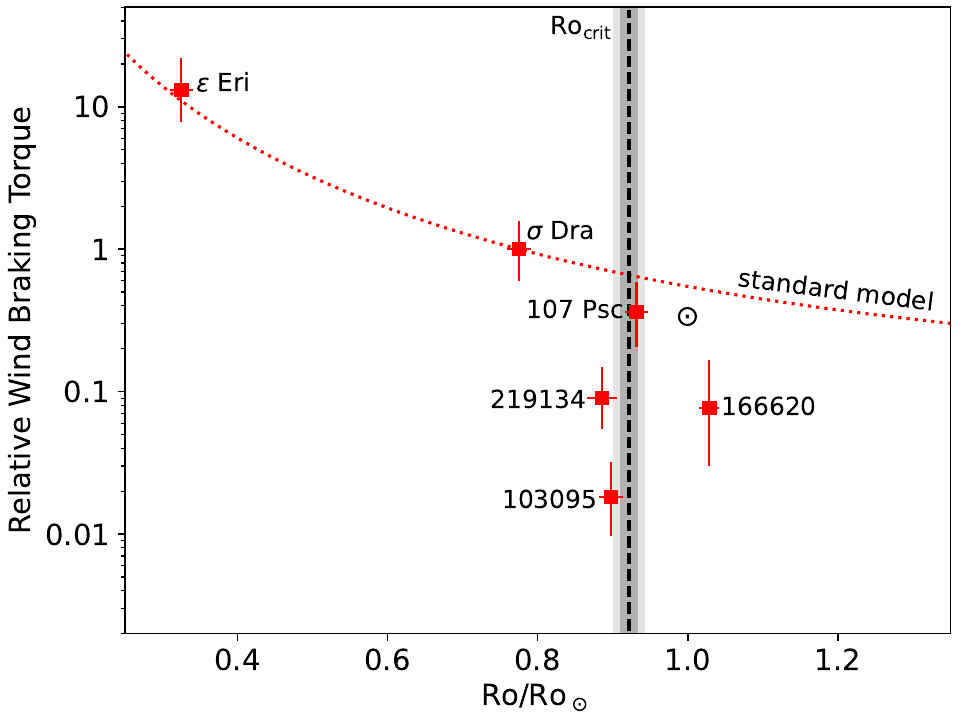}
 \caption{Estimated wind braking torque relative to \sigDra as a function of Ro 
normalized to the solar value. The light gray shaded area indicates the systematic 
uncertainty in Ro/$\Rosun$, while the darker gray shaded area marks the empirical onset 
of WMB in solar analogs. The solar point $\odot$ is from \cite{Finley2018}.\label{fig5}}
 \end{figure}

In Section~\ref{sec2} we obtained constraints on most of the inputs that are required to 
calculate the wind braking torque using the prescription of 
\cite{FinleyMatt2018}\footnote{\url{https://github.com/travismetcalfe/FinleyMatt2018}}. 
Spectropolarimetry provided constraints on the equivalent polar field strengths for the 
large-scale magnetic field ($B_{\rm d}, B_{\rm q}, B_{\rm o}$), archival \xray 
measurements produced estimates of mass-loss rates from the empirical relation of 
\cite{Wood2021}, and SED fitting yielded stellar radii and masses. In this section we 
combine these inputs with rotation periods measured from the Mount Wilson survey 
\citep{Baliunas1996}, and from \cite{Folsom2018b} for HD\,219134, to estimate the wind 
braking torque for each of our targets. We also evaluate the relative importance of 
various contributions to the overall decrease in the torque along the evolutionary 
sequence. The resulting estimates of wind braking torque are listed in Table~\ref{tab1} 
and illustrated in Figure~\ref{fig5}, with uncertainties defined by simultaneously 
shifting all of the inputs to their $\pm1\sigma$ values to minimize or maximize the 
torque.

For previous samples of hotter stars, we adopted the asteroseismic calibration of 
convective overturn times from \cite{Corsaro2021} to calculate Rossby numbers. 
Unfortunately, this relation is only calibrated for Gaia colors between $0.55 < 
G_{BP}-G_{RP} < 0.97$, and all of the stars in our K dwarf sample are redder. 
\cite{Metcalfe2024c} recently extended the relation to redder colors, using measured 
rotation periods and mean activity levels from the Mount Wilson survey to estimate the 
convective overturn times. The results suggested that the observed deviation from 
linearity in \cite{Corsaro2021} at $G_{BP}-G_{RP} > 0.85$ was probably an observational 
bias against the detection of solar-like oscillations in more active K-type stars. The 
extended relation is nearly linear between $0.55 < G_{BP}-G_{RP} < 1.2$, and allows us to 
estimate Ro/$\Rosun$ from the inverse of the mean activity level relative to the solar 
value from \cite{Egeland2017}. For consistency with previous results, we use this 
formulation to estimate Ro/$\Rosun$ for the early K-type stars in our sample.

To establish a context for our estimated wind braking torques, Figure~\ref{fig5} includes 
a standard spin-down model for \sigDra (red dotted line), as well as the empirical value 
of $\Rocrit$ for the onset of WMB in solar analogs \citep{Metcalfe2022}. The targets with 
Ro/$\Rosun$ well below $\Rocrit$ are consistent with the standard spin-down 
model\footnote{The standard model is from a fit to \sigDra, so the agreement with \epsEri 
may be fortuitous. The scatter in the empirical relation of \cite{Wood2021} is large at 
high activity levels, up to two orders of magnitude.}, while those near and above 
$\Rocrit$ show varying degrees of WMB (discussed below). The most evolved star in 
Figure~\ref{fig5}---the magnetic grand minimum star HD\,166620---is well beyond $\Rocrit$ 
and has an estimated wind braking torque that is substantially below the standard 
spin-down predictions. This pattern broadly corroborates our findings from similar wind 
braking estimates for hotter stars \citep{Metcalfe2021, Metcalfe2022, Metcalfe2023a, 
Metcalfe2024a}.

Our wind braking estimate for \epsEri adopts a direct inference of the mass-loss rate 
from Ly$\alpha$ measurements. The mass-loss rate implied by the \xray luminosity and 
stellar radius listed in Table~\ref{tab1} is somewhat lower than the direct inference 
($9.9^{+10.5}_{-5.6}\ \dot{M}_\odot$), which would yield a weaker torque 
($7.6^{+5.1}_{-3.3} \times 10^{30}$~erg) near the lower bound in Figure~\ref{fig5}. 
Considering the lower metallicity of \sigDra, the resulting deviation from the standard 
spin-down model would not fundamentally alter our interpretation \citep[e.g., 
see][]{Amard2020}.

The most evolved target that has not yet entered the WMB regime is \sigDra, with a wind 
braking torque that is more than an order of magnitude weaker than \epsEri ($-92\%$). To 
assess the relative importance of various contributions to the overall decrease in the 
wind braking torque, we can modify one parameter at a time between the fiducial models 
for these two stars. We assume that magnetic field strength scales with the observed 
$\meanRhk$ and is reflected in the absolute values of ($B_{\rm d}, B_{\rm q}, B_{\rm 
o}$), while the field morphology is reflected in their relative values. As expected for 
standard spin-down, the total decrease in wind braking torque is dominated by 
evolutionary changes in the mass-loss rate ($-71\%$), magnetic field strength and 
morphology ($-58\%$), and rotation period ($-56\%$). These contributions are slightly 
offset by differences in the stellar radius ($+40\%$) and mass ($+0.5\%$). For this pair 
of stars, the combined influence of field strength and morphology is dominated by 
evolutionary changes in magnetic field strength ($-55\%$) with a very small contribution 
from differences in morphology ($-7\%$).

The youngest target that may have recently entered the WMB regime is \Psc, with a wind 
braking torque several times weaker than \sigDra ($-64\%$) falling slightly below the 
standard spin-down prediction. Comparing the fiducial models for these two stars, the 
total decrease in wind braking torque is once again dominated by evolutionary changes in 
the mass-loss rate ($-47\%$), magnetic field strength and morphology ($-23\%$), and 
rotation period ($-23\%$), with a small contribution from the difference in stellar mass 
($-0.8\%$) offset by the difference in radius ($+18\%$). In this case, the combined 
influence of field strength and morphology is still primarily from evolutionary changes 
in magnetic field strength ($-16\%$), but with a comparable contribution due to changes 
in morphology ($-9\%$).

Our most evolved target with a ZDI map is HD\,219134, which has a wind braking torque 
more than an order of magnitude weaker than \sigDra ($-91\%$), well within the WMB 
regime. For this pair of stars, the total decrease in wind braking torque is still 
dominated by evolutionary changes in the mass-loss rate ($-62\%$), magnetic field 
strength and morphology ($-55\%$), and rotation period ($-36\%$). These contributions are 
reinforced in this case by differences in the stellar radius ($-18\%$), and slightly 
offset by a small difference in mass ($+1.1\%$). In contrast to our evolutionary sequence 
in the standard spin-down regime, the combined influence of field strength and morphology 
is now dominated by differences in morphology ($-49\%$) with a much smaller contribution 
from differences in magnetic field strength ($-12\%$). Like \epsEri, the fiducial model 
for HD\,219134 adopts a direct inference of the mass-loss rate from Ly$\alpha$ 
measurements. The mass-loss rate implied by the \xray luminosity and stellar radius 
listed in Table~\ref{tab1} is higher ($0.75^{+0.75}_{-0.38}\ \dot{M}_\odot$), which would 
yield a slightly stronger torque ($0.119^{+0.073}_{-0.046} \times 10^{30}$~erg) that 
would shift the point upwards in Figure~\ref{fig5} by roughly its own size. This would 
reduce the relative importance of changes in the mass-loss rate (from $-62\%$ to $-53\%$) 
without altering our other conclusions.

For the final two targets in our sample (HD\,103095 and HD\,166620) we have made the 
conservative assumption that all of the large-scale field is in an axisymmetric dipole 
configuration. This prevents us from drawing firm conclusions about changes in the 
magnetic morphology relative to field strength, but we can still assess the importance of 
other changes. The estimated wind braking torque for HD\,103095 is more than fifty times 
weaker than \sigDra ($-98\%$), a decrease that might be enhanced by the extreme low 
metallicity of this star (as suggested previously by metal-poor $\tau$\,Cet). In this 
case, the total decrease in wind braking torque is dominated by evolutionary changes in 
the magnetic field strength and morphology ($-88\%$), mass-loss rate ($-72\%$), and 
rotation period ($-13\%$), reinforced by differences in stellar radius ($-44\%$) and 
offset by differences in the mass ($+8\%$).

The oldest star in our sample is HD\,166620, with a wind braking torque that is more than 
an order of magnitude weaker than \sigDra ($-92\%$). As with HD\,103095, the total 
decrease in wind braking torque is dominated by evolutionary changes in the magnetic 
field strength and morphology ($-78\%$), mass-loss rate ($-48\%$), and rotation period 
($-37\%$). These contributions are slightly offset by differences in the stellar radius 
($+4\%$) and mass ($+1.1\%$). For both HD\,103095 and HD\,166620, note that the combined 
influence of field strength and morphology may be even more important if the magnetic 
field is actually comprised of a mixture of dipole, quadrupole, and octupole components.

\section{Summary and Discussion}\label{sec4}

We have assembled new and archival observations (Section~\ref{sec2}) to estimate the wind 
braking torque for an evolutionary sequence of six early K-type stars using the 
prescription of \cite{FinleyMatt2018}\footnote{The referee suggests that, because 
the underlying torque model used a polytropic Parker wind profile that is known to be an 
inaccurate representation of the physics driving the solar wind, application of this 
approach across a broad range of activity levels may be subject to systematic errors.}. 
The resulting constraints on magnetic braking (Table~\ref{tab1}) 
complement our earlier applications to two late F-type stars \citep{Metcalfe2021}, five 
solar analogs \citep{Metcalfe2022, Metcalfe2024a}, and two late G-type stars 
\citep{Metcalfe2023a}, gradually extending the approach to the more challenging cooler 
targets. The evolutionary picture that emerges (Section~\ref{sec3}) is broadly consistent 
with our previous findings, revealing a dramatic transition from the relatively well 
calibrated standard spin-down behavior in more active stars like \epsEri and \sigDra to 
substantially weakened braking in the least active targets like the magnetic grand 
minimum star HD\,166620 (Figure~\ref{fig5}).

As with the hotter targets, the wind braking torque in early K-type stars decreases by 
roughly an order of magnitude as they reach a critical value of the Rossby number 
($\Rocrit \sim 0.9\ \Rosun$). We observe a drop in both magnetic field strength and \xray 
luminosity across this boundary. The onset of WMB occurs at the same $\Rocrit$ as the 
threshold for hotter stars, even though the characteristic rotation periods for 
post-transition stars are significantly longer here (\mbox{31--43~days}), compared to 
21--23~days for solar analogs \citep{Metcalfe2022, Metcalfe2024a} and even shorter for 
late F-type stars \citep{Metcalfe2019, Metcalfe2021}. Our finding confirms the Rossby 
number as the key global stellar predictor of WMB.

The larger number of targets, and the longer evolutionary timescales of K dwarfs, have 
allowed us to probe the onset of WMB in greater detail than we could in the hotter stars. 
For the first time, we may have caught one star (\Psc) on the cusp of the transition from 
standard spin-down to WMB, showing a moderately reduced braking torque driven primarily 
by weaker mass-loss with a smaller contribution from magnetic properties and rotation. In 
the standard spin-down regime (from \epsEri to \sigDra) the contribution of magnetic 
properties to the evolution of wind braking torque can be attributed almost entirely to 
changes in the field strength ($-55\%$), with very little change in magnetic morphology 
($-7\%$). By contrast, in the WMB regime (from \sigDra to HD\,219134) the magnetic 
contributions are almost reversed, with a large decrease in wind braking torque due to a 
magnetic morphology shift ($-49\%$) and very little change due to field strength 
($-12\%$). Understanding how this apparent trend continues to evolve within the WMB 
regime will require detailed ZDI mapping of older targets such as HD\,166620, which is in 
a prolonged flat activity phase and shows extremely small variations in circular 
polarization.

The ubiquity of stellar activity cycles in this sample is a natural consequence of the 
longer evolutionary timescales of K dwarfs and the finite age of the Galaxy. In a similar 
sample of solar analogs, we found one star in the WMB regime with an activity cycle 
\citep[18\,Sco;][]{Metcalfe2022}, while older stars in that sample (16\,Cyg\,A and 
16\,Cyg\,B) showed constant activity over decades \citep{Radick2018}. This is consistent 
with previous suggestions that the onset of WMB corresponds to the disruption of 
large-scale organization by the stellar dynamo, with activity cycles growing longer and 
weaker over roughly the second half of main-sequence lifetimes \citep{Metcalfe2017}. In 
contrast to our hotter targets, there are no observations of ``flat activity'' among the 
K stars \citep{Baliunas1995}. The only apparent exception is the K-type subgiant 
$\delta$\,Eri, which originated as an F-type star on the main-sequence 
\citep{Carrier2003}.

The fact that the WMB regime only begins after $\sim$7~Gyr in K dwarfs, combined with 
their longer main-sequence lifetimes, implies that even the oldest stars in our sample 
are still in the early phases of this magnetic transition. As a consequence, activity 
cycles are still evident in the K dwarfs that exhibit WMB---with the exception of 
HD\,166620, which transitioned from cycling to flat activity near the end of the Mount 
Wilson survey \citep{Baum2022, Luhn2022}. If the value of $\Rocrit$ for the onset of WMB 
is also the critical value for the efficient operation of a large-scale dynamo 
\citep{DurneyLatour1978, Metcalfe2020, Metcalfe2024b}, then the activity cycles in K 
dwarfs older than $\sim$7~Gyr may represent subcritical stellar dynamos 
\citep{Tripathi2021}.

Future observations promise to improve our characterization of the transition to WMB in 
this sample of early K-type stars. Direct inferences of the mass-loss rate for \sigDra 
and HD\,166620 may be possible from Ly$\alpha$ measurements, and the required Hubble 
Space Telescope observations have been scheduled for May 2025 (HST-GO-17793, PI: 
B.~Wood). Even in the absence of a mass-loss detection for HD\,166620, an updated \xray 
luminosity may soon emerge from a recent 34~ks XMM observation (PI: B.~Stelzer), which 
can be compared to our ROSAT-era measurement while it was still cycling in the 1990s. 
Time series spectropolarimetry of HD\,166620 for the construction of a ZDI map will be 
challenging---both because of the tiny Stokes~$V$ variations (e.g., Figure~\ref{fig1}), 
and the longer rotation period that requires a sustained observing campaign---but the 
effort is justified by the relatively unexplored domain at high Rossby number that it 
seems to occupy.

\vspace*{12pt}
The authors would like to thank Tim Brown, Ricky Egeland, and Brian Wood for helpful 
exchanges. T.S.M.\ acknowledges support from NSF grant AST-2205919 and NASA grant 
80NSSC22K0475. Computational time at the Texas Advanced Computing Center was provided 
through allocation TG-AST090107. J.v.S.\ acknowledges support from NSF grant AST-2205888. 
O.K.\ acknowledges support by the Swedish Research Council (grant agreements no.\ 
2019-03548 and 2023-03667). K.G.S.\ acknowledges funding for PEPSI 
(\url{https://pepsi.aip.de/}) through the German Verbundforschung grants 05AL2BA1/3 and 
05A08BAC as well as the State of Brandenburg. A.J.F.\ acknowledges support from the 
European Research Council (ERC) under the European Union's Horizon 2020 research and 
innovation program (grant agreement No.\ 810218 WHOLESUN). R.A.G.\ acknowledges support 
from the PLATO and GOLF Centre National D'{\'{E}}tudes Spatiales grants. The LBT is an 
international collaboration among institutions in the United States, Italy and Germany. 
LBT Corporation partners are: The University of Arizona on behalf of the Arizona Board of 
Regents; Istituto Nazionale di Astrofisica, Italy; LBT Beteiligungsgesellschaft, Germany, 
representing the Max-Planck Society, The Leibniz Institute for Astrophysics Potsdam, and 
Heidelberg University; The Ohio State University, and The Research Corporation, on behalf 
of The University of Notre Dame, University of Minnesota and University of Virginia. This 
paper includes data collected with the TESS mission, obtained from the Mikulski Archive 
for Space Telescopes at the Space Telescope Science Institute. The specific observations 
analyzed can be accessed via 
\dataset[doi:10.17909/x1b4-qh75]{https://doi.org/10.17909/x1b4-qh75}. Funding for the 
TESS mission is provided by the NASA Explorer Program. 

\appendix\vspace*{-24pt}
\section{Asteroseismic Nondetections from TESS}\label{appa}

\begin{figure}[b!]
 \centering\includegraphics[width=5.0in]{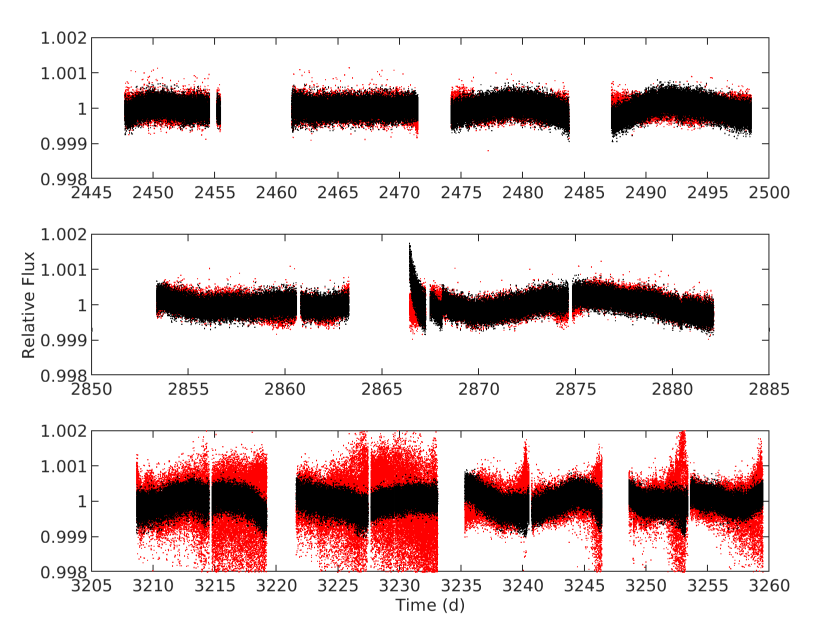}
 \caption{Extracted light curves for \Psc from TESS sectors 42 and 43 (top), 57 (middle), 
and 70 and 71 (bottom). Results from the custom pipeline used here are shown in black, 
while those from the standard SPOC pipeline are in red. Our pipeline results are shown 
prior to high-pass filtering. For the first three sectors shown, the quality of the two 
light curves are substantially similar, but when the scattered light level rises in 
Sectors 70 and 71, our custom pipeline yields roughly twice the S/N of the SPOC 
product.\label{fig6}}
 \end{figure}

We extracted TESS light curves for our targets using the custom approach described in 
\cite{Nielsen2020} and \cite{Metcalfe2023b}. This approach involves the creation of 
aperture masks which optimize the high-frequency S/N in the light curve by growing the 
aperture one pixel at a time. We then detrended the light curve against the centroid 
position and high-pass filtered using a cutoff frequency of $100~\mu$Hz. The quality of 
the resulting light curves ranges from only nominal improvement over the Science 
Processing Operations Center (SPOC) products to as much as a factor of two reduction in 
noise. Figure~\ref{fig6} shows this entire range in the light curve for one target, \Psc. 
In the earlier sectors where the scattered light levels are relatively low, our approach 
leads to roughly similar outcomes, but in later sectors where scattered light levels rise 
substantially, our custom approach does approximately a factor of two better, yielding a 
noticeably improved S/N in the amplitude spectrum.

We analyzed the custom light curves for \Psc, HD\,219134 and HD\,166620 to search for 
solar-like oscillations using pySYD \citep{Chontos2022}, yielding null detections in all 
cases. To derive upper limits on the oscillation amplitudes, we evaluated the background 
model from pySYD at the predicted frequency of maximum power $\nu_{\rm max}$ for each 
target \citep{Hey2024}. We required a height-to-background ratio of 1.1 
\citep{Mosser2012}, which is typically sufficient for a detection. The resulting upper 
limits were 16.5~ppm for \Psc (at~3583~$\mu$Hz), 10.6~ppm for HD\,219134 
(at~4751~$\mu$Hz), and 15.2~ppm for HD\,166620 (at~3814~$\mu$Hz). Upper limits for 
\epsEri and HD\,103095 were previously published in \cite{Metcalfe2023b}.

We attempted to improve on the detection of solar-like oscillations in \sigDra published 
by \cite{Hon2024}, extracting a custom light curve and incorporating additional 
observations through Sector~80. The resulting light curve had a S/N about 10\% higher 
than the SPOC product, and we tentatively identified several quadrupole modes to help 
constrain the stellar age. We identified a total of 23 oscillation modes using a Bayesian 
peak-bagging package \citep{Breton2022} and used the Asteroseismic Modeling Portal v2.0 
\citep[AMP;][]{AMP2023} for updated modeling. However, the results did not differ 
significantly from those published in \cite{Hon2024}, so we adopted the latter for the 
stellar properties listed in Table~\ref{tab1}.

\begin{thebibliography}{}
\expandafter\ifx\csname natexlab\endcsname\relax\def\natexlab#1{#1}\fi
\providecommand{\url}[1]{\href{#1}{#1}}
\providecommand{\dodoi}[1]{doi:~\href{http://doi.org/#1}{\nolinkurl{#1}}}
\providecommand{\doeprint}[1]{\href{http://ascl.net/#1}{\nolinkurl{http://ascl.net/#1}}}
\providecommand{\doarXiv}[1]{\href{https://arxiv.org/abs/#1}{\nolinkurl{https://arxiv.org/abs/#1}}}

\bibitem[{{Amard} {et~al.}(2020){Amard}, {Roquette}, \& {Matt}}]{Amard2020}
{Amard}, L., {Roquette}, J., \& {Matt}, S.~P. 2020, \mnras, 499, 3481,
  \dodoi{10.1093/mnras/staa3038}

\bibitem[{{Auri{\`e}re}(2003)}]{Auriere2003}
{Auri{\`e}re}, M. 2003, in EAS Publications Series, Vol.~9, EAS Publications
  Series, ed. J.~{Arnaud} \& N.~{Meunier}, 105

\bibitem[{{Ayres}(2024)}]{Ayres2024}
{Ayres}, T. 2024, \aj, submitted

\bibitem[{{Baliunas} {et~al.}(1996){Baliunas}, {Sokoloff}, \&
  {Soon}}]{Baliunas1996}
{Baliunas}, S., {Sokoloff}, D., \& {Soon}, W. 1996, \apjl, 457, L99,
  \dodoi{10.1086/309891}

\bibitem[{{Baliunas} {et~al.}(1995){Baliunas}, {Donahue}, {Soon}, {Horne},
  {Frazer}, {Woodard-Eklund}, {Bradford}, {Rao}, {Wilson}, {Zhang}, {Bennett},
  {Briggs}, {Carroll}, {Duncan}, {Figueroa}, {Lanning}, {Misch}, {Mueller},
  {Noyes}, {Poppe}, {Porter}, {Robinson}, {Russell}, {Shelton}, {Soyumer},
  {Vaughan}, \& {Whitney}}]{Baliunas1995}
{Baliunas}, S.~L., {Donahue}, R.~A., {Soon}, W.~H., {et~al.} 1995, \apj, 438,
  269, \dodoi{10.1086/175072}

\bibitem[{{Basu} \& {Antia}(1995)}]{Basu1995}
{Basu}, S., \& {Antia}, H.~M. 1995, \mnras, 276, 1402,
  \dodoi{10.1093/mnras/276.4.1402}

\bibitem[{{Baum} {et~al.}(2022){Baum}, {Wright}, {Luhn}, \&
  {Isaacson}}]{Baum2022}
{Baum}, A.~C., {Wright}, J.~T., {Luhn}, J.~K., \& {Isaacson}, H. 2022, \aj,
  163, 183, \dodoi{10.3847/1538-3881/ac5683}

\bibitem[{{Bellotti} {et~al.}(2024){Bellotti}, {Morin}, {Lehmann}, {Petit},
  {Hussain}, {Donati}, {Folsom}, {Carmona}, {Martioli}, {Klein}, {Fouqu{\'e}},
  {Moutou}, {Alencar}, {Artigau}, {Boisse}, {Bouchy}, {Bouvier}, {Cook},
  {Delfosse}, {Doyon}, \& {H{\'e}brard}}]{bellotti2024}
{Bellotti}, S., {Morin}, J., {Lehmann}, L.~T., {et~al.} 2024, \aap, 686, A66,
  \dodoi{10.1051/0004-6361/202348043}

\bibitem[{{Boro Saikia} {et~al.}(2018){Boro Saikia}, {Lueftinger}, {Jeffers},
  {Folsom}, {See}, {Petit}, {Marsden}, {Vidotto}, {Morin}, {Reiners}, {Guedel},
  \& {BCool Collaboration}}]{BoroSaikia2018}
{Boro Saikia}, S., {Lueftinger}, T., {Jeffers}, S.~V., {et~al.} 2018, \aap,
  620, L11, \dodoi{10.1051/0004-6361/201834347}

\bibitem[{{Bowler} {et~al.}(2023){Bowler}, {Tran}, {Zhang}, {Morgan}, {Ashok},
  {Blunt}, {Bryan}, {Evans}, {Franson}, {Huber}, {Nagpal}, {Wu}, \&
  {Zhou}}]{Bowler2023}
{Bowler}, B.~P., {Tran}, Q.~H., {Zhang}, Z., {et~al.} 2023, \aj, 165, 164,
  \dodoi{10.3847/1538-3881/acbd34}

\bibitem[{{Breton} {et~al.}(2022){Breton}, {Garc{\'\i}a}, {Ballot}, {Delsanti},
  \& {Salabert}}]{Breton2022}
{Breton}, S.~N., {Garc{\'\i}a}, R.~A., {Ballot}, J., {Delsanti}, V., \&
  {Salabert}, D. 2022, \aap, 663, A118, \dodoi{10.1051/0004-6361/202243330}

\bibitem[{{Brewer} {et~al.}(2016){Brewer}, {Fischer}, {Valenti}, \&
  {Piskunov}}]{Brewer2016}
{Brewer}, J.~M., {Fischer}, D.~A., {Valenti}, J.~A., \& {Piskunov}, N. 2016,
  \apjs, 225, 32, \dodoi{10.3847/0067-0049/225/2/32}

\bibitem[{{Brown}(2014)}]{Brown2014}
{Brown}, T.~M. 2014, \apj, 789, 101, \dodoi{10.1088/0004-637X/789/2/101}

\bibitem[{{Brun} {et~al.}(2022){Brun}, {Strugarek}, {Noraz}, {Perri}, {Varela},
  {Augustson}, {Charbonneau}, \& {Toomre}}]{Brun2022}
{Brun}, A.~S., {Strugarek}, A., {Noraz}, Q., {et~al.} 2022, \apj, 926, 21,
  \dodoi{10.3847/1538-4357/ac469b}

\bibitem[{{Buzasi}(1997)}]{Buzasi1997}
{Buzasi}, D.~L. 1997, \apj, 484, 855, \dodoi{10.1086/304374}

\bibitem[{{Cao} \& {Pinsonneault}(2022)}]{Cao2022}
{Cao}, L., \& {Pinsonneault}, M.~H. 2022, \mnras, 517, 2165,
  \dodoi{10.1093/mnras/stac2706}

\bibitem[{{Carrier} {et~al.}(2003){Carrier}, {Bouchy}, \&
  {Eggenberger}}]{Carrier2003}
{Carrier}, F., {Bouchy}, F., \& {Eggenberger}, P. 2003, in Asteroseismology
  Across the HR Diagram, ed. M.~J. {Thompson}, M.~S. {Cunha}, \& M.~J.~P.~F.~G.
  {Monteiro}, Vol. 284, 311--314

\bibitem[{{Chontos} {et~al.}(2022){Chontos}, {Huber}, {Sayeed}, \&
  {Yamsiri}}]{Chontos2022}
{Chontos}, A., {Huber}, D., {Sayeed}, M., \& {Yamsiri}, P. 2022, The Journal of
  Open Source Software, 7, 3331, \dodoi{10.21105/joss.03331}

\bibitem[{{Coffaro} {et~al.}(2020){Coffaro}, {Stelzer}, {Orlando}, {Hall},
  {Metcalfe}, {Wolter}, {Mittag}, {Sanz-Forcada}, {Schneider}, \&
  {Ducci}}]{Coffaro2020}
{Coffaro}, M., {Stelzer}, B., {Orlando}, S., {et~al.} 2020, \aap, 636, A49,
  \dodoi{10.1051/0004-6361/201936479}

\bibitem[{{Corsaro} {et~al.}(2021){Corsaro}, {Bonanno}, {Mathur},
  {Garc{\'\i}a}, {Santos}, {Breton}, \& {Khalatyan}}]{Corsaro2021}
{Corsaro}, E., {Bonanno}, A., {Mathur}, S., {et~al.} 2021, \aap, 652, L2,
  \dodoi{10.1051/0004-6361/202141395}

\bibitem[{{Curtis} {et~al.}(2019){Curtis}, {Ag{\"u}eros}, {Douglas}, \&
  {Meibom}}]{Curtis2019}
{Curtis}, J.~L., {Ag{\"u}eros}, M.~A., {Douglas}, S.~T., \& {Meibom}, S. 2019,
  \apj, 879, 49, \dodoi{10.3847/1538-4357/ab2393}

\bibitem[{{Curtis} {et~al.}(2020){Curtis}, {Ag{\"u}eros}, {Matt}, {Covey},
  {Douglas}, {Angus}, {Saar}, {Cody}, {Vanderburg}, {Law}, {Kraus}, {Latham},
  {Baranec}, {Riddle}, {Ziegler}, {Lund}, {Torres}, {Meibom}, {Aguirre}, \&
  {Wright}}]{Curtis2020}
{Curtis}, J.~L., {Ag{\"u}eros}, M.~A., {Matt}, S.~P., {et~al.} 2020, \apj, 904,
  140, \dodoi{10.3847/1538-4357/abbf58}

\bibitem[{{David} {et~al.}(2022){David}, {Angus}, {Curtis}, {van Saders},
  {Colman}, {Contardo}, {Lu}, \& {Zinn}}]{David2022}
{David}, T.~J., {Angus}, R., {Curtis}, J.~L., {et~al.} 2022, \apj, 933, 114,
  \dodoi{10.3847/1538-4357/ac6dd3}

\bibitem[{{do Nascimento} {et~al.}(2023){do Nascimento}, {Barnes}, {Saar}, {de
  Mello}, {Hall}, {Anthony}, {de Almeida}, {Velloso}, {da Costa}, {Petit},
  {Strugarek}, {Wargelin}, {Castro}, {Strassmeier}, \&
  {Brun}}]{doNascimento2023}
{do Nascimento}, J.~D., {Barnes}, S.~A., {Saar}, S.~H., {et~al.} 2023, \apj,
  958, 57, \dodoi{10.3847/1538-4357/acfc1a}

\bibitem[{{Donati} {et~al.}(1997){Donati}, {Semel}, {Carter}, {Rees}, \&
  {Collier Cameron}}]{Donati1997}
{Donati}, J.~F., {Semel}, M., {Carter}, B.~D., {Rees}, D.~E., \& {Collier
  Cameron}, A. 1997, \mnras, 291, 658, \dodoi{10.1093/mnras/291.4.658}

\bibitem[{{Donati} {et~al.}(2006){Donati}, {Howarth}, {Jardine}, {Petit},
  {Catala}, {Landstreet}, {Bouret}, {Alecian}, {Barnes}, {Forveille},
  {Paletou}, \& {Manset}}]{Donati2006}
{Donati}, J.~F., {Howarth}, I.~D., {Jardine}, M.~M., {et~al.} 2006, \mnras,
  370, 629, \dodoi{10.1111/j.1365-2966.2006.10558.x}

\bibitem[{{Dotter} {et~al.}(2017){Dotter}, {Conroy}, {Cargile}, \&
  {Asplund}}]{Dotter2017}
{Dotter}, A., {Conroy}, C., {Cargile}, P., \& {Asplund}, M. 2017, \apj, 840,
  99, \dodoi{10.3847/1538-4357/aa6d10}

\bibitem[{{Durney} \& {Latour}(1978)}]{DurneyLatour1978}
{Durney}, B.~R., \& {Latour}, J. 1978, Geophysical and Astrophysical Fluid
  Dynamics, 9, 241, \dodoi{10.1080/03091927708242330}

\bibitem[{{Egeland} {et~al.}(2017){Egeland}, {Soon}, {Baliunas}, {Hall},
  {Pevtsov}, \& {Bertello}}]{Egeland2017}
{Egeland}, R., {Soon}, W., {Baliunas}, S., {et~al.} 2017, \apj, 835, 25,
  \dodoi{10.3847/1538-4357/835/1/25}

\bibitem[{{Epstein} \& {Pinsonneault}(2014)}]{Epstein2014}
{Epstein}, C.~R., \& {Pinsonneault}, M.~H. 2014, \apj, 780, 159,
  \dodoi{10.1088/0004-637X/780/2/159}

\bibitem[{{Finley} \& {Matt}(2018)}]{FinleyMatt2018}
{Finley}, A.~J., \& {Matt}, S.~P. 2018, \apj, 854, 78,
  \dodoi{10.3847/1538-4357/aaaab5}

\bibitem[{{Finley} {et~al.}(2018){Finley}, {Matt}, \& {See}}]{Finley2018}
{Finley}, A.~J., {Matt}, S.~P., \& {See}, V. 2018, \apj, 864, 125,
  \dodoi{10.3847/1538-4357/aad7b6}

\bibitem[{{Folsom} {et~al.}(2018{\natexlab{a}}){Folsom}, {Fossati}, {Wood},
  {Sreejith}, {Cubillos}, {Vidotto}, {Alecian}, {Girish}, {Lichtenegger},
  {Murthy}, {Petit}, \& {Valyavin}}]{Folsom2018b}
{Folsom}, C.~P., {Fossati}, L., {Wood}, B.~E., {et~al.} 2018{\natexlab{a}},
  \mnras, 481, 5286, \dodoi{10.1093/mnras/sty2494}

\bibitem[{{Folsom} {et~al.}(2018{\natexlab{b}}){Folsom}, {Bouvier}, {Petit},
  {L{\`e}bre}, {Amard}, {Palacios}, {Morin}, {Donati}, \&
  {Vidotto}}]{Folsom2018a}
{Folsom}, C.~P., {Bouvier}, J., {Petit}, P., {et~al.} 2018{\natexlab{b}},
  \mnras, 474, 4956, \dodoi{10.1093/mnras/stx3021}

\bibitem[{{Fuhrmeister} {et~al.}(2023){Fuhrmeister}, {Coffaro}, {Stelzer},
  {Mittag}, {Czesla}, \& {Schneider}}]{Fuhrmeister2023}
{Fuhrmeister}, B., {Coffaro}, M., {Stelzer}, B., {et~al.} 2023, \aap, 672,
  A149, \dodoi{10.1051/0004-6361/202245201}

\bibitem[{{Gaia Collaboration} {et~al.}(2021){Gaia Collaboration}, {Brown},
  {Vallenari}, {Prusti}, {de Bruijne}, {Babusiaux}, {Biermann}, {Creevey},
  {Evans}, {Eyer}, {Hutton}, {Jansen}, {Jordi}, {Klioner}, {Lammers},
  {Lindegren}, {Luri}, {Mignard}, {Panem}, {Pourbaix}, {Randich}, {Sartoretti},
  {Soubiran}, {Walton}, {Arenou}, {Bailer-Jones}, {Bastian}, {Cropper},
  {Drimmel}, {Katz}, {Lattanzi}, {van Leeuwen}, {Bakker}, {Cacciari},
  {Casta{\~n}eda}, {De Angeli}, {Ducourant}, {Fabricius}, {Fouesneau},
  {Fr{\'e}mat}, {Guerra}, {Guerrier}, {Guiraud}, {Jean-Antoine Piccolo},
  {Masana}, {Messineo}, {Mowlavi}, {Nicolas}, {Nienartowicz}, {Pailler},
  {Panuzzo}, {Riclet}, {Roux}, {Seabroke}, {Sordo}, {Tanga}, {Th{\'e}venin},
  {Gracia-Abril}, {Portell}, {Teyssier}, {Altmann}, {Andrae}, {Bellas-Velidis},
  {Benson}, {Berthier}, {Blomme}, {Brugaletta}, {Burgess}, {Busso}, {Carry},
  {Cellino}, {Cheek}, {Clementini}, {Damerdji}, {Davidson}, {Delchambre},
  {Dell'Oro}, {Fern{\'a}ndez-Hern{\'a}ndez}, {Galluccio}, {Garc{\'\i}a-Lario},
  {Garcia-Reinaldos}, {Gonz{\'a}lez-N{\'u}{\~n}ez}, {Gosset}, {Haigron},
  {Halbwachs}, {Hambly}, {Harrison}, {Hatzidimitriou}, {Heiter},
  {Hern{\'a}ndez}, {Hestroffer}, {Hodgkin}, {Holl}, {Jan{\ss}en}, {Jevardat de
  Fombelle}, {Jordan}, {Krone-Martins}, {Lanzafame}, {L{\"o}ffler}, {Lorca},
  {Manteiga}, {Marchal}, {Marrese}, {Moitinho}, {Mora}, {Muinonen}, {Osborne},
  {Pancino}, {Pauwels}, {Petit}, {Recio-Blanco}, {Richards}, {Riello},
  {Rimoldini}, {Robin}, {Roegiers}, {Rybizki}, {Sarro}, {Siopis}, {Smith},
  {Sozzetti}, {Ulla}, {Utrilla}, {van Leeuwen}, {van Reeven}, {Abbas}, {Abreu
  Aramburu}, {Accart}, {Aerts}, {Aguado}, {Ajaj}, {Altavilla}, {{\'A}lvarez},
  {{\'A}lvarez Cid-Fuentes}, {Alves}, {Anderson}, {Anglada Varela}, {Antoja},
  {Audard}, {Baines}, {Baker}, {Balaguer-N{\'u}{\~n}ez}, {Balbinot}, {Balog},
  {Barache}, {Barbato}, {Barros}, {Barstow}, {Bartolom{\'e}}, {Bassilana},
  {Bauchet}, {Baudesson-Stella}, {Becciani}, {Bellazzini}, {Bernet}, {Bertone},
  {Bianchi}, {Blanco-Cuaresma}, {Boch}, {Bombrun}, {Bossini}, {Bouquillon},
  {Bragaglia}, {Bramante}, {Breedt}, {Bressan}, {Brouillet}, {Bucciarelli},
  {Burlacu}, {Busonero}, {Butkevich}, {Buzzi}, {Caffau}, {Cancelliere},
  {C{\'a}novas}, {Cantat-Gaudin}, {Carballo}, {Carlucci}, {Carnerero},
  {Carrasco}, {Casamiquela}, {Castellani}, {Castro-Ginard}, {Castro Sampol},
  {Chaoul}, {Charlot}, {Chemin}, {Chiavassa}, {Cioni}, {Comoretto}, {Cooper},
  {Cornez}, {Cowell}, {Crifo}, {Crosta}, {Crowley}, {Dafonte}, {Dapergolas},
  {David}, {David}, {de Laverny}, {De Luise}, {De March}, {De Ridder}, {de
  Souza}, {de Teodoro}, {de Torres}, {del Peloso}, {del Pozo}, {Delbo},
  {Delgado}, {Delgado}, {Delisle}, {Di Matteo}, {Diakite}, {Diener},
  {Distefano}, {Dolding}, {Eappachen}, {Edvardsson}, {Enke}, {Esquej}, {Fabre},
  {Fabrizio}, {Faigler}, {Fedorets}, {Fernique}, {Fienga}, {Figueras},
  {Fouron}, {Fragkoudi}, {Fraile}, {Franke}, {Gai}, {Garabato},
  {Garcia-Gutierrez}, {Garc{\'\i}a-Torres}, {Garofalo}, {Gavras}, {Gerlach},
  {Geyer}, {Giacobbe}, {Gilmore}, {Girona}, {Giuffrida}, {Gomel}, {Gomez},
  {Gonzalez-Santamaria}, {Gonz{\'a}lez-Vidal}, {Granvik},
  {Guti{\'e}rrez-S{\'a}nchez}, {Guy}, {Hauser}, {Haywood}, {Helmi}, {Hidalgo},
  {Hilger}, {H{\l}adczuk}, {Hobbs}, {Holland}, {Huckle}, {Jasniewicz},
  {Jonker}, {Juaristi Campillo}, {Julbe}, {Karbevska}, {Kervella}, {Khanna},
  {Kochoska}, {Kontizas}, {Kordopatis}, {Korn}, {Kostrzewa-Rutkowska},
  {Kruszy{\'n}ska}, {Lambert}, {Lanza}, {Lasne}, {Le Campion}, {Le Fustec},
  {Lebreton}, {Lebzelter}, {Leccia}, {Leclerc}, {Lecoeur-Taibi}, {Liao},
  {Licata}, {Lindstr{\o}m}, {Lister}, {Livanou}, {Lobel}, {Madrero Pardo},
  {Managau}, {Mann}, {Marchant}, {Marconi}, {Marcos Santos}, {Marinoni},
  {Marocco}, {Marshall}, {Martin Polo}, {Mart{\'\i}n-Fleitas}, {Masip},
  {Massari}, {Mastrobuono-Battisti}, {Mazeh}, {McMillan}, {Messina},
  {Michalik}, {Millar}, {Mints}, {Molina}, {Molinaro}, {Moln{\'a}r},
  {Montegriffo}, {Mor}, {Morbidelli}, {Morel}, {Morris}, {Mulone}, {Munoz},
  {Muraveva}, {Murphy}, {Musella}, {Noval}, {Ord{\'e}novic}, {Orr{\`u}},
  {Osinde}, {Pagani}, {Pagano}, {Palaversa}, {Palicio}, {Panahi}, {Pawlak},
  {Pe{\~n}alosa Esteller}, {Penttil{\"a}}, {Piersimoni}, {Pineau}, {Plachy},
  {Plum}, {Poggio}, {Poretti}, {Poujoulet}, {Pr{\v{s}}a}, {Pulone}, {Racero},
  {Ragaini}, {Rainer}, {Raiteri}, {Rambaux}, {Ramos}, {Ramos-Lerate}, {Re
  Fiorentin}, {Regibo}, {Reyl{\'e}}, {Ripepi}, {Riva}, {Rixon}, {Robichon},
  {Robin}, {Roelens}, {Rohrbasser}, {Romero-G{\'o}mez}, {Rowell}, {Royer},
  {Rybicki}, {Sadowski}, {Sagrist{\`a} Sell{\'e}s}, {Sahlmann}, {Salgado},
  {Salguero}, {Samaras}, {Sanchez Gimenez}, {Sanna}, {Santove{\~n}a},
  {Sarasso}, {Schultheis}, {Sciacca}, {Segol}, {Segovia}, {S{\'e}gransan},
  {Semeux}, {Shahaf}, {Siddiqui}, {Siebert}, {Siltala}, {Slezak}, {Smart},
  {Solano}, {Solitro}, {Souami}, {Souchay}, {Spagna}, {Spoto}, {Steele},
  {Steidelm{\"u}ller}, {Stephenson}, {S{\"u}veges}, {Szabados}, {Szegedi-Elek},
  {Taris}, {Tauran}, {Taylor}, {Teixeira}, {Thuillot}, {Tonello}, {Torra},
  {Torra}, {Turon}, {Unger}, {Vaillant}, {van Dillen}, {Vanel}, {Vecchiato},
  {Viala}, {Vicente}, {Voutsinas}, {Weiler}, {Wevers}, {Wyrzykowski}, {Yoldas},
  {Yvard}, {Zhao}, {Zorec}, {Zucker}, {Zurbach}, \& {Zwitter}}]{Gaia2021}
{Gaia Collaboration}, {Brown}, A.~G.~A., {Vallenari}, A., {et~al.} 2021, \aap,
  649, A1, \dodoi{10.1051/0004-6361/202039657}

\bibitem[{{Garc{\'\i}a} {et~al.}(2014){Garc{\'\i}a}, {Ceillier}, {Salabert},
  {Mathur}, {van Saders}, {Pinsonneault}, {Ballot}, {Beck}, {Bloemen},
  {Campante}, {Davies}, {do Nascimento}, {Mathis}, {Metcalfe}, {Nielsen},
  {Su{\'a}rez}, {Chaplin}, {Jim{\'e}nez}, \& {Karoff}}]{Garcia2014}
{Garc{\'\i}a}, R.~A., {Ceillier}, T., {Salabert}, D., {et~al.} 2014, \aap, 572,
  A34, \dodoi{10.1051/0004-6361/201423888}

\bibitem[{{Garraffo} {et~al.}(2016){Garraffo}, {Drake}, \&
  {Cohen}}]{Garraffo2016}
{Garraffo}, C., {Drake}, J.~J., \& {Cohen}, O. 2016, \aap, 595, A110,
  \dodoi{10.1051/0004-6361/201628367}

\bibitem[{{Garraffo} {et~al.}(2018){Garraffo}, {Drake}, {Dotter}, {Choi},
  {Burke}, {Moschou}, {Alvarado-G{\'o}mez}, {Kashyap}, \&
  {Cohen}}]{Garraffo2018}
{Garraffo}, C., {Drake}, J.~J., {Dotter}, A., {et~al.} 2018, \apj, 862, 90,
  \dodoi{10.3847/1538-4357/aace5d}

\bibitem[{{Hall} {et~al.}(2021){Hall}, {Davies}, {van Saders}, {Nielsen},
  {Lund}, {Chaplin}, {Garc{\'\i}a}, {Amard}, {Breimann}, {Khan}, {See}, \&
  {Tayar}}]{Hall2021}
{Hall}, O.~J., {Davies}, G.~R., {van Saders}, J., {et~al.} 2021, Nature
  Astronomy, 5, 707, \dodoi{10.1038/s41550-021-01335-x}

\bibitem[{{Hey} {et~al.}(2024){Hey}, {Huber}, {Ong}, {Stello}, \&
  {Foreman-Mackey}}]{Hey2024}
{Hey}, D., {Huber}, D., {Ong}, J., {Stello}, D., \& {Foreman-Mackey}, D. 2024,
  \apjs, submitted (arXiv:2403.02489), \dodoi{10.48550/arXiv.2403.02489}

\bibitem[{{Hon} {et~al.}(2024){Hon}, {Huber}, {Li}, {Metcalfe}, {Bedding},
  {Ong}, {Chontos}, {Rubenzahl}, {Halverson}, {Garc{\'\i}a}, {Kjeldsen},
  {Stello}, {Hey}, {Campante}, {Howard}, {Gibson}, {Rider}, {Roy}, {Baker},
  {Edelstein}, {Smith}, {Fulton}, {Walawender}, {Brodheim}, {Brown}, {Chan},
  {Dai}, {Deich}, {Gottschalk}, {Grillo}, {Hale}, {Hill}, {Holden},
  {Householder}, {Isaacson}, {Ishikawa}, {Jelinsky}, {Kassis}, {Kaye}, {Laher},
  {Lanclos}, {Lee}, {Lilley}, {McCarney}, {Miller}, {Payne}, {Petigura},
  {Poppett}, {Raffanti}, {Rockosi}, {Sanford}, {Schwab}, {Shaum}, {Sirk},
  {Smith}, {Thorne}, {Valliant}, {Vandenberg}, {Wang}, {Wishnow}, {Wold},
  {Yeh}, {Baker}, {Basu}, {Bedell}, {Cegla}, {Crossfield}, {Dressing},
  {Dumusque}, {Knutson}, {Mawet}, {O'Meara}, {Stef{\'a}nsson}, {Teske},
  {Vasisht}, {Wang}, {Weiss}, {Winn}, \& {Wright}}]{Hon2024}
{Hon}, M., {Huber}, D., {Li}, Y., {et~al.} 2024, \apj, 975, 147,
  \dodoi{10.3847/1538-4357/ad76a9}

\bibitem[{{Jeffers} {et~al.}(2014){Jeffers}, {Petit}, {Marsden}, {Morin},
  {Donati}, \& {Folsom}}]{Jeffers2014}
{Jeffers}, S.~V., {Petit}, P., {Marsden}, S.~C., {et~al.} 2014, \aap, 569, A79,
  \dodoi{10.1051/0004-6361/201423725}

\bibitem[{{Johnson} {et~al.}(2016){Johnson}, {Endl}, {Cochran}, {Meschiari},
  {Robertson}, {MacQueen}, {Brugamyer}, {Caldwell}, {Hatzes}, {Ram{\'\i}rez},
  \& {Wittenmyer}}]{Johnson2016}
{Johnson}, M.~C., {Endl}, M., {Cochran}, W.~D., {et~al.} 2016, \apj, 821, 74,
  \dodoi{10.3847/0004-637X/821/2/74}

\bibitem[{{Kawaler}(1988)}]{Kawaler1988}
{Kawaler}, S.~D. 1988, \apj, 333, 236, \dodoi{10.1086/166740}

\bibitem[{{Kochukhov} {et~al.}(2010){Kochukhov}, {Makaganiuk}, \&
  {Piskunov}}]{Kochukhov2010}
{Kochukhov}, O., {Makaganiuk}, V., \& {Piskunov}, N. 2010, \aap, 524, A5,
  \dodoi{10.1051/0004-6361/201015429}

\bibitem[{{Kurucz}(1993)}]{Kurucz1993}
{Kurucz}, R.~L. 1993, in Astronomical Society of the Pacific Conference Series,
  Vol.~44, IAU Colloq. 138: Peculiar versus Normal Phenomena in A-type and
  Related Stars, ed. M.~M. {Dworetsky}, F.~{Castelli}, \& R.~{Faraggiana}, 87

\bibitem[{{Li} {et~al.}(2025){Li}, {Huber}, {Ong}, {van Saders}, {Costa},
  {Reersted Larsen}, {Basu}, {Bedding}, {Dai}, {Chontos}, {Carmichael}, {Hey},
  {Kjeldsen}, {Hon}, {Campante}, {Monteiro}, {Sloth Lundkvist}, {Saunders},
  {Isaacson}, {Howard}, {Gibson}, {Halverson}, {Rider}, {Roy}, {Baker},
  {Edelstein}, {Smith}, {Fulton}, \& {Walawender}}]{Li2025}
{Li}, Y., {Huber}, D., {Ong}, J.~M.~J., {et~al.} 2025, \apj, submitted,
  (arXiv:2502.00971), \dodoi{10.48550/arXiv.2502.00971}

\bibitem[{{Lu} {et~al.}(2024){Lu}, {Angus}, {Foreman-Mackey}, \&
  {Hattori}}]{Lu2024}
{Lu}, Y., {Angus}, R., {Foreman-Mackey}, D., \& {Hattori}, S. 2024, \aj, 167,
  159, \dodoi{10.3847/1538-3881/ad28b9}

\bibitem[{{Lu} {et~al.}(2021){Lu}, {Angus}, {Curtis}, {David}, \&
  {Kiman}}]{Lu2021}
{Lu}, Y.~L., {Angus}, R., {Curtis}, J.~L., {David}, T.~J., \& {Kiman}, R. 2021,
  \aj, 161, 189, \dodoi{10.3847/1538-3881/abe4d6}

\bibitem[{{Lu} {et~al.}(2022){Lu}, {Curtis}, {Angus}, {David}, \&
  {Hattori}}]{Lu2022}
{Lu}, Y.~L., {Curtis}, J.~L., {Angus}, R., {David}, T.~J., \& {Hattori}, S.
  2022, \aj, 164, 251, \dodoi{10.3847/1538-3881/ac9bee}

\bibitem[{{Luhn} {et~al.}(2022){Luhn}, {Wright}, {Henry}, {Saar}, \&
  {Baum}}]{Luhn2022}
{Luhn}, J.~K., {Wright}, J.~T., {Henry}, G.~W., {Saar}, S.~H., \& {Baum}, A.~C.
  2022, \apjl, 936, L23, \dodoi{10.3847/2041-8213/ac8b13}

\bibitem[{{Masuda}(2022)}]{Masuda2022}
{Masuda}, K. 2022, \apj, 937, 94, \dodoi{10.3847/1538-4357/ac8d58}

\bibitem[{{McQuillan} {et~al.}(2014){McQuillan}, {Mazeh}, \&
  {Aigrain}}]{McQuillan2014}
{McQuillan}, A., {Mazeh}, T., \& {Aigrain}, S. 2014, \apjs, 211, 24,
  \dodoi{10.1088/0067-0049/211/2/24}

\bibitem[{{Mermilliod}(2006)}]{Mermilliod2006}
{Mermilliod}, J.~C. 2006, VizieR Online Data Catalog, II/168

\bibitem[{{Metcalfe} {et~al.}(2024{\natexlab{a}}){Metcalfe}, {Corsaro},
  {Bonanno}, {Creevey}, \& {van Saders}}]{Metcalfe2024c}
{Metcalfe}, T.~S., {Corsaro}, E., {Bonanno}, A., {Creevey}, O.~L., \& {van
  Saders}, J.~L. 2024{\natexlab{a}}, RNAAS, 8, 260,
  \dodoi{10.3847/2515-5172/ad8566}

\bibitem[{{Metcalfe} {et~al.}(2019){Metcalfe}, {Kochukhov}, {Ilyin},
  {Strassmeier}, {Godoy-Rivera}, \& {Pinsonneault}}]{Metcalfe2019}
{Metcalfe}, T.~S., {Kochukhov}, O., {Ilyin}, I.~V., {et~al.} 2019, \apjl, 887,
  L38, \dodoi{10.3847/2041-8213/ab5e48}

\bibitem[{{Metcalfe} {et~al.}(2023{\natexlab{a}}){Metcalfe}, {Townsend}, \&
  {Ball}}]{AMP2023}
{Metcalfe}, T.~S., {Townsend}, R. H.~D., \& {Ball}, W.~H. 2023{\natexlab{a}},
  RNAAS, 7, 164, \dodoi{10.3847/2515-5172/acebef}

\bibitem[{{Metcalfe} \& {van Saders}(2017)}]{Metcalfe2017}
{Metcalfe}, T.~S., \& {van Saders}, J. 2017, \solphys, 292, 126,
  \dodoi{10.1007/s11207-017-1157-5}

\bibitem[{{Metcalfe} {et~al.}(2013){Metcalfe}, {Buccino}, {Brown}, {Mathur},
  {Soderblom}, {Henry}, {Mauas}, {Petrucci}, {Hall}, \& {Basu}}]{Metcalfe2013}
{Metcalfe}, T.~S., {Buccino}, A.~P., {Brown}, B.~P., {et~al.} 2013, \apjl, 763,
  L26, \dodoi{10.1088/2041-8205/763/2/L26}

\bibitem[{{Metcalfe} {et~al.}(2014){Metcalfe}, {Creevey}, {Do{\u g}an},
  {Mathur}, {Xu}, {Bedding}, {Chaplin}, {Christensen-Dalsgaard}, {Karoff},
  {Trampedach}, {Benomar}, {Brown}, {Buzasi}, {Campante}, {{\c C}elik},
  {Cunha}, {Davies}, {Deheuvels}, {Derekas}, {Di Mauro}, {Garc{\'{\i}}a},
  {Guzik}, {Howe}, {MacGregor}, {Mazumdar}, {Montalb{\'a}n}, {Monteiro},
  {Salabert}, {Serenelli}, {Stello}, {Ste{\c s}licki}, {Suran}, {Y{\i}ld{\i}z},
  {Aksoy}, {Elsworth}, {Gruberbauer}, {Guenther}, {Lebreton}, {Molaverdikhani},
  {Pricopi}, {Simoniello}, \& {White}}]{Metcalfe2014}
{Metcalfe}, T.~S., {Creevey}, O.~L., {Do{\u g}an}, G., {et~al.} 2014, \apjs,
  214, 27, \dodoi{10.1088/0067-0049/214/2/27}

\bibitem[{{Metcalfe} {et~al.}(2020){Metcalfe}, {van Saders}, {Basu}, {Buzasi},
  {Chaplin}, {Egeland}, {Garcia}, {Gaulme}, {Huber}, {Reinhold}, {Schunker},
  {Stassun}, {Appourchaux}, {Ball}, {Bedding}, {Deheuvels},
  {Gonz{\'a}lez-Cuesta}, {Handberg}, {Jim{\'e}nez}, {Kjeldsen}, {Li}, {Lund},
  {Mathur}, {Mosser}, {Nielsen}, {Noll}, {{\c{C}}elik Orhan}, {{\"O}rtel},
  {Santos}, {Yildiz}, {Baliunas}, \& {Soon}}]{Metcalfe2020}
{Metcalfe}, T.~S., {van Saders}, J.~L., {Basu}, S., {et~al.} 2020, \apj, 900,
  154, \dodoi{10.3847/1538-4357/aba963}

\bibitem[{{Metcalfe} {et~al.}(2021){Metcalfe}, {van Saders}, {Basu}, {Buzasi},
  {Drake}, {Egeland}, {Huber}, {Saar}, {Stassun}, {Ball}, {Campante}, {Finley},
  {Kochukhov}, {Mathur}, {Reinhold}, {See}, {Baliunas}, \&
  {Soon}}]{Metcalfe2021}
---. 2021, \apj, 921, 122, \dodoi{10.3847/1538-4357/ac1f19}

\bibitem[{{Metcalfe} {et~al.}(2022){Metcalfe}, {Finley}, {Kochukhov}, {See},
  {Ayres}, {Stassun}, {van Saders}, {Clark}, {Godoy-Rivera}, {Ilyin},
  {Pinsonneault}, {Strassmeier}, \& {Petit}}]{Metcalfe2022}
{Metcalfe}, T.~S., {Finley}, A.~J., {Kochukhov}, O., {et~al.} 2022, \apjl, 933,
  L17, \dodoi{10.3847/2041-8213/ac794d}

\bibitem[{{Metcalfe} {et~al.}(2023{\natexlab{b}}){Metcalfe}, {Strassmeier},
  {Ilyin}, {van Saders}, {Ayres}, {Finley}, {Kochukhov}, {Petit}, {See},
  {Stassun}, {Jeffers}, {Marsden}, {Morin}, \& {Vidotto}}]{Metcalfe2023a}
{Metcalfe}, T.~S., {Strassmeier}, K.~G., {Ilyin}, I.~V., {et~al.}
  2023{\natexlab{b}}, \apjl, 948, L6, \dodoi{10.3847/2041-8213/acce38}

\bibitem[{{Metcalfe} {et~al.}(2023{\natexlab{c}}){Metcalfe}, {Buzasi}, {Huber},
  {Pinsonneault}, {van Saders}, {Ayres}, {Basu}, {Drake}, {Egeland},
  {Kochukhov}, {Petit}, {Saar}, {See}, {Stassun}, {Li}, {Bedding}, {Breton},
  {Finley}, {Garcia}, {Kjeldsen}, {Nielsen}, {Ong}, {Rorsted}, {Stokholm},
  {Winther}, {Clark}, {Godoy-Rivera}, {Ilyin}, {Strassmeier}, {Jeffers},
  {Marsden}, {Vidotto}, {Baliunas}, \& {Soon}}]{Metcalfe2023b}
{Metcalfe}, T.~S., {Buzasi}, D., {Huber}, D., {et~al.} 2023{\natexlab{c}}, \aj,
  166, 167, \dodoi{10.3847/1538-3881/acf1f7}

\bibitem[{{Metcalfe} {et~al.}(2024{\natexlab{b}}){Metcalfe}, {Strassmeier},
  {Ilyin}, {Buzasi}, {Kochukhov}, {Ayres}, {Basu}, {Chontos}, {Finley}, {See},
  {Stassun}, {van Saders}, {Sepulveda}, \& {Ricker}}]{Metcalfe2024a}
{Metcalfe}, T.~S., {Strassmeier}, K.~G., {Ilyin}, I.~V., {et~al.}
  2024{\natexlab{b}}, \apjl, 960, L6, \dodoi{10.3847/2041-8213/ad0a95}

\bibitem[{{Metcalfe} {et~al.}(2024{\natexlab{c}}){Metcalfe}, {van Saders},
  {Huber}, {Buzasi}, {Garc{\'\i}a}, {Stassun}, {Basu}, {Breton}, {Claytor},
  {Corsaro}, {Nielsen}, {Ong}, {Saunders}, {Stokholm}, \&
  {Bedding}}]{Metcalfe2024b}
{Metcalfe}, T.~S., {van Saders}, J.~L., {Huber}, D., {et~al.}
  2024{\natexlab{c}}, \apj, 974, 31, \dodoi{10.3847/1538-4357/ad6dd6}

\bibitem[{{Mosser} {et~al.}(2012){Mosser}, {Elsworth}, {Hekker}, {Huber},
  {Kallinger}, {Mathur}, {Belkacem}, {Goupil}, {Samadi}, {Barban}, {Bedding},
  {Chaplin}, {Garc{\'\i}a}, {Stello}, {De Ridder}, {Middour}, {Morris}, \&
  {Quintana}}]{Mosser2012}
{Mosser}, B., {Elsworth}, Y., {Hekker}, S., {et~al.} 2012, \aap, 537, A30,
  \dodoi{10.1051/0004-6361/201117352}

\bibitem[{{Nielsen} {et~al.}(2020){Nielsen}, {Ball}, {Standing}, {Triaud},
  {Buzasi}, {Carboneau}, {Stassun}, {Kane}, {Chaplin}, {Bellinger}, {Mosser},
  {Roxburgh}, {{\c{C}}elik Orhan}, {Y{\i}ld{\i}z}, {{\"O}rtel}, {Vrard},
  {Mazumdar}, {Ranadive}, {Deal}, {Davies}, {Campante}, {Garc{\'\i}a},
  {Mathur}, {Gonz{\'a}lez-Cuesta}, \& {Serenelli}}]{Nielsen2020}
{Nielsen}, M.~B., {Ball}, W.~H., {Standing}, M.~R., {et~al.} 2020, \aap, 641,
  A25, \dodoi{10.1051/0004-6361/202037461}

\bibitem[{{Noyes} {et~al.}(1984){Noyes}, {Hartmann}, {Baliunas}, {Duncan}, \&
  {Vaughan}}]{Noyes1984}
{Noyes}, R.~W., {Hartmann}, L.~W., {Baliunas}, S.~L., {Duncan}, D.~K., \&
  {Vaughan}, A.~H. 1984, \apj, 279, 763, \dodoi{10.1086/161945}

\bibitem[{{Paunzen}(2015)}]{Paunzen2015}
{Paunzen}, E. 2015, \aap, 580, A23, \dodoi{10.1051/0004-6361/201526413}

\bibitem[{{Petit} {et~al.}(2014){Petit}, {Louge}, {Th{\'e}ado}, {Paletou},
  {Manset}, {Morin}, {Marsden}, \& {Jeffers}}]{Petit2014}
{Petit}, P., {Louge}, T., {Th{\'e}ado}, S., {et~al.} 2014, \pasp, 126, 469,
  \dodoi{10.1086/676976}

\bibitem[{{Petit} {et~al.}(2021){Petit}, {Folsom}, {Donati}, {Yu}, {do
  Nascimento}, {Jeffers}, {Marsden}, {Morin}, \& {Vidotto}}]{Petit2021}
{Petit}, P., {Folsom}, C.~P., {Donati}, J.~F., {et~al.} 2021, \aap, 648, A55,
  \dodoi{10.1051/0004-6361/202040027}

\bibitem[{{Pinsonneault} {et~al.}(1989){Pinsonneault}, {Kawaler}, {Sofia}, \&
  {Demarque}}]{Pinsonneault1989}
{Pinsonneault}, M.~H., {Kawaler}, S.~D., {Sofia}, S., \& {Demarque}, P. 1989,
  \apj, 338, 424, \dodoi{10.1086/167210}

\bibitem[{{Radick} {et~al.}(2018){Radick}, {Lockwood}, {Henry}, {Hall}, \&
  {Pevtsov}}]{Radick2018}
{Radick}, R.~R., {Lockwood}, G.~W., {Henry}, G.~W., {Hall}, J.~C., \&
  {Pevtsov}, A.~A. 2018, \apj, 855, 75, \dodoi{10.3847/1538-4357/aaaae3}

\bibitem[{{Reiners} {et~al.}(2022){Reiners}, {Shulyak}, {K{\"a}pyl{\"a}},
  {Ribas}, {Nagel}, {Zechmeister}, {Caballero}, {Shan}, {Fuhrmeister},
  {Quirrenbach}, {Amado}, {Montes}, {Jeffers}, {Azzaro}, {B{\'e}jar},
  {Chaturvedi}, {Henning}, {K{\"u}rster}, \& {Pall{\'e}}}]{Reiners2022}
{Reiners}, A., {Shulyak}, D., {K{\"a}pyl{\"a}}, P.~J., {et~al.} 2022, \aap,
  662, A41, \dodoi{10.1051/0004-6361/202243251}

\bibitem[{{Ryabchikova} {et~al.}(2015){Ryabchikova}, {Piskunov}, {Kurucz},
  {Stempels}, {Heiter}, {Pakhomov}, \& {Barklem}}]{Ryabchikova2015}
{Ryabchikova}, T., {Piskunov}, N., {Kurucz}, R.~L., {et~al.} 2015, \physscr,
  90, 054005, \dodoi{10.1088/0031-8949/90/5/054005}

\bibitem[{{Santos} {et~al.}(2019){Santos}, {Garc{\'\i}a}, {Mathur}, {Bugnet},
  {van Saders}, {Metcalfe}, {Simonian}, \& {Pinsonneault}}]{Santos2019}
{Santos}, A.~R.~G., {Garc{\'\i}a}, R.~A., {Mathur}, S., {et~al.} 2019, \apjs,
  244, 21, \dodoi{10.3847/1538-4365/ab3b56}

\bibitem[{{Saunders} {et~al.}(2024){Saunders}, {van Saders}, {Lyttle},
  {Metcalfe}, {Li}, {Davies}, {Hall}, {Ball}, {Townsend}, {Creevey}, \&
  {Dodds}}]{Saunders2024}
{Saunders}, N., {van Saders}, J.~L., {Lyttle}, A.~J., {et~al.} 2024, \apj, 962,
  138, \dodoi{10.3847/1538-4357/ad1516}

\bibitem[{{See} {et~al.}(2016){See}, {Jardine}, {Vidotto}, {Donati}, {Boro
  Saikia}, {Bouvier}, {Fares}, {Folsom}, {Gregory}, {Hussain}, {Jeffers},
  {Marsden}, {Morin}, {Moutou}, {do Nascimento}, {Petit}, \& {Waite}}]{see2016}
{See}, V., {Jardine}, M., {Vidotto}, A.~A., {et~al.} 2016, \mnras, 462, 4442,
  \dodoi{10.1093/mnras/stw2010}

\bibitem[{{Semel}(1989)}]{Semel1989}
{Semel}, M. 1989, \aap, 225, 456

\bibitem[{{Skumanich}(1972)}]{Skumanich1972}
{Skumanich}, A. 1972, \apj, 171, 565, \dodoi{10.1086/151310}

\bibitem[{{Spada} \& {Lanzafame}(2020)}]{Spada2020}
{Spada}, F., \& {Lanzafame}, A.~C. 2020, \aap, 636, A76,
  \dodoi{10.1051/0004-6361/201936384}

\bibitem[{{Stassun} {et~al.}(2017){Stassun}, {Collins}, \&
  {Gaudi}}]{Stassun2017}
{Stassun}, K.~G., {Collins}, K.~A., \& {Gaudi}, B.~S. 2017, \aj, 153, 136,
  \dodoi{10.3847/1538-3881/aa5df3}

\bibitem[{{Stassun} {et~al.}(2018){Stassun}, {Corsaro}, {Pepper}, \&
  {Gaudi}}]{Stassun2018}
{Stassun}, K.~G., {Corsaro}, E., {Pepper}, J.~A., \& {Gaudi}, B.~S. 2018, \aj,
  155, 22, \dodoi{10.3847/1538-3881/aa998a}

\bibitem[{{Stassun} \& {Torres}(2016)}]{Stassun2016}
{Stassun}, K.~G., \& {Torres}, G. 2016, \apjl, 831, L6,
  \dodoi{10.3847/2041-8205/831/1/L6}

\bibitem[{{Stassun} \& {Torres}(2021)}]{StassunTorres2021}
---. 2021, \apjl, 907, L33, \dodoi{10.3847/2041-8213/abdaad}

\bibitem[{{Strassmeier} {et~al.}(2015){Strassmeier}, {Ilyin}, {J{\"a}rvinen},
  {Weber}, {Woche}, {Barnes}, {Bauer}, {Beckert}, {Bittner}, {Bredthauer},
  {Carroll}, {Denker}, {Dionies}, {DiVarano}, {D{\"o}scher}, {Fechner},
  {Feuerstein}, {Granzer}, {Hahn}, {Harnisch}, {Hofmann}, {Lesser}, {Paschke},
  {Pankratow}, {Plank}, {Pl{\"u}schke}, {Popow}, \&
  {Sablowski}}]{Strassmeier2015}
{Strassmeier}, K.~G., {Ilyin}, I., {J{\"a}rvinen}, A., {et~al.} 2015,
  Astronomische Nachrichten, 336, 324, \dodoi{10.1002/asna.201512172}

\bibitem[{{Suzuki}(2013)}]{Suzuki2013}
{Suzuki}, T.~K. 2013, Astronomische Nachrichten, 334, 81,
  \dodoi{10.1002/asna.201211751}

\bibitem[{{Tayar} {et~al.}(2022){Tayar}, {Claytor}, {Huber}, \& {van
  Saders}}]{Tayar2022}
{Tayar}, J., {Claytor}, Z.~R., {Huber}, D., \& {van Saders}, J. 2022, \apj,
  927, 31, \dodoi{10.3847/1538-4357/ac4bbc}

\bibitem[{{Thoul} {et~al.}(1994){Thoul}, {Bahcall}, \& {Loeb}}]{Thoul1994}
{Thoul}, A.~A., {Bahcall}, J.~N., \& {Loeb}, A. 1994, \apj, 421, 828,
  \dodoi{10.1086/173695}

\bibitem[{{Torres} {et~al.}(2010){Torres}, {Andersen}, \&
  {Gim{\'e}nez}}]{Torres2010}
{Torres}, G., {Andersen}, J., \& {Gim{\'e}nez}, A. 2010, \aapr, 18, 67,
  \dodoi{10.1007/s00159-009-0025-1}

\bibitem[{{Tripathi} {et~al.}(2021){Tripathi}, {Nandy}, \&
  {Banerjee}}]{Tripathi2021}
{Tripathi}, B., {Nandy}, D., \& {Banerjee}, S. 2021, \mnras, 506, L50,
  \dodoi{10.1093/mnrasl/slab035}

\bibitem[{{Valenti} \& {Fischer}(2005)}]{ValentiFischer2005}
{Valenti}, J.~A., \& {Fischer}, D.~A. 2005, \apjs, 159, 141,
  \dodoi{10.1086/430500}

\bibitem[{{van~Saders} {et~al.}(2016){van~Saders}, {Ceillier}, {Metcalfe},
  {Silva Aguirre}, {Pinsonneault}, {Garc{\'\i}a}, {Mathur}, \&
  {Davies}}]{vanSaders2016}
{van~Saders}, J.~L., {Ceillier}, T., {Metcalfe}, T.~S., {et~al.} 2016, \nat,
  529, 181, \dodoi{10.1038/nature16168}

\bibitem[{{van~Saders} \& {Pinsonneault}(2013)}]{vanSaders2013}
{van~Saders}, J.~L., \& {Pinsonneault}, M.~H. 2013, \apj, 776, 67,
  \dodoi{10.1088/0004-637X/776/2/67}

\bibitem[{{van~Saders} {et~al.}(2019){van~Saders}, {Pinsonneault}, \&
  {Barbieri}}]{vanSaders2019}
{van~Saders}, J.~L., {Pinsonneault}, M.~H., \& {Barbieri}, M. 2019, \apj, 872,
  128, \dodoi{10.3847/1538-4357/aafafe}

\bibitem[{{Vidotto} {et~al.}(2014){Vidotto}, {Gregory}, {Jardine}, {Donati},
  {Petit}, {Morin}, {Folsom}, {Bouvier}, {Cameron}, {Hussain}, {Marsden},
  {Waite}, {Fares}, {Jeffers}, \& {do Nascimento}}]{vidotto2014}
{Vidotto}, A.~A., {Gregory}, S.~G., {Jardine}, M., {et~al.} 2014, \mnras, 441,
  2361, \dodoi{10.1093/mnras/stu728}

\bibitem[{{Weber} \& {Davis}(1967)}]{WeberDavis1967}
{Weber}, E.~J., \& {Davis}, Leverett, J. 1967, \apj, 148, 217,
  \dodoi{10.1086/149138}

\bibitem[{{Wood}(2018)}]{Wood2018}
{Wood}, B.~E. 2018, in Journal of Physics Conference Series, Vol. 1100, Journal
  of Physics Conference Series, 012028

\bibitem[{{Wood} {et~al.}(2005b){Wood}, {M{\"u}ller}, {Zank}, {Linsky}, \&
  {Redfield}}]{Wood2005a}
{Wood}, B.~E., {M{\"u}ller}, H.~R., {Zank}, G.~P., {Linsky}, J.~L., \&
  {Redfield}, S. 2005b, \apjl, 628, L143, \dodoi{10.1086/432716}

\bibitem[{{Wood} {et~al.}(2021){Wood}, {M{\"u}ller}, {Redfield}, {Konow},
  {Vannier}, {Linsky}, {Youngblood}, {Vidotto}, {Jardine},
  {Alvarado-G{\'o}mez}, \& {Drake}}]{Wood2021}
{Wood}, B.~E., {M{\"u}ller}, H.-R., {Redfield}, S., {et~al.} 2021, \apj, 915,
  37, \dodoi{10.3847/1538-4357/abfda5}

\end{thebibliography}
\end{document}